\documentclass[12pt, draftclsnofoot, onecolumn]{IEEEtran}
\pdfoutput=1
\usepackage{color}
\usepackage{cite}
\usepackage[final]{graphicx}
\usepackage{amsmath}
\usepackage{amssymb}
\usepackage{amsfonts}
\usepackage{array}
\usepackage{stfloats}
\usepackage{lineno}
\usepackage{booktabs}
\usepackage{tabularx}
\usepackage{caption}
\usepackage{epstopdf}
\usepackage{amsthm}
\usepackage{multirow}
\usepackage{multicol}
\usepackage{lipsum}
\usepackage[ruled]{algorithm2e}
\usepackage{setspace}
\usepackage{soul}
\usepackage[english]{babel}
\usepackage{subcaption}
\usepackage{color}
\usepackage{colortbl}

\newcolumntype{C}{>{\centering\arraybackslash$}p{\linewidth}<{$}}

\newtheorem{remark}{Remark}
\newtheorem{theorem}{Theorem}

\newtheorem{lemma}{Lemma}

\newtheorem{corollary}{Corollary}

\newcommand{\RNum}[1]{\uppercase\expandafter{\romannumeral #1\relax}}

\begin{document}
	
	\title{Joint Location and Beamforming Design for STAR-RIS Assisted NOMA Systems}
	
	\author{Qiling~Gao, Yuanwei~Liu,~\IEEEmembership{Senior Member,~IEEE}, Xidong~Mu,~\IEEEmembership{Graduate Student Member,~IEEE}, Min~Jia,~\IEEEmembership{Senior Member,~IEEE}, Dongbo~Li, and~Lajos~Hanzo,~\IEEEmembership{Life~Fellow,~IEEE}
		\thanks{Q. Gao, M. Jia and D. Li are with Harbin Institute of Technology, Harbin, 150008, China. (email: 18B905013@stu.hit.edu.cn,~jiamin@hit.edu.cn,~ldb@hit.edu.cn).	
			X. Mu and Y. Liu are with Queen Mary University of London, London, UK. (email: x.mu@qmul.ac.uk,~yuanwei.liu@qmul.ac.uk).
			%
			L. Hanzo is with the School of Electronics and Computer Science, University of Southampton, Southampton, SO17 1BJ, U.K. (e-mail: lh@ecs.soton.ac.uk).
		}
	}
	
	\maketitle
	\vspace{-40pt}
	\begin{abstract}
		Simultaneously transmitting and reflecting reconfigurable intelligent surface (STAR-RIS) assisted non-orthogonal multiple access (NOMA) communication systems are investigated in its vicinity, where a STAR-RIS is deployed within a predefined region for establishing communication links for users. Both beamformer-based NOMA and cluster-based NOMA schemes are employed at the multi-antenna base station (BS). For each scheme, the STAR-RIS deployment location, the passive transmitting and reflecting beamforming (BF) of the STAR-RIS, and the active BF at the BS are jointly optimized for maximizing the weighted sum-rate (WSR) of users. To solve the resultant non-convex problems, an alternating optimization (AO) algorithm is proposed, where successive convex approximation (SCA) and semi-definite programming (SDP) methods are invoked for iteratively addressing the non-convexity of each sub-problem. Numerical results reveal that 1) the WSR performance can be significantly enhanced by optimizing the specific deployment location of the STAR-RIS; 2) both beamformer-based and cluster-based NOMA prefer asymmetric STAR-RIS deployment.
		
	\end{abstract}
	
	\vspace{-15pt}
	\begin{IEEEkeywords}
		Beamforming design, deployment design, multi-antenna NOMA, STAR-RIS.
	\end{IEEEkeywords}

	%
	\IEEEpeerreviewmaketitle
	\vspace{-15pt}
	\section{Introduction}
	
	Next-generation wireless networks are expected to cope with the explosive proliferation of wireless devices in a spectral- and energy-efficiency manner, which requires the development of revolutionary techniques\cite{saad2019vision,yang20196g,zhang20196g}. Among others, as a planar meta-surface having a large number of reconfigurable passive elements, reconfigurable intelligent surfaces (RIS) constitute a promising technology, which is capable beneficially ameliorating the propagation of the incident signal by adjusting the phase and amplitude of each element, hence facilitating the creation of conducive environments \cite{elzanaty2021reconfigurable,9324795,zhang2020reconfigurable,DiRenzoSmart2020}. However, conventional RISs are only capable of reflecting signals, hence only users roaming in the 180$ ^\circ $ half-plane can
	deployed at the same side of RISs with respect to the transmitters can be served \cite{mu2021joint}. 
	As a remedy, a novel category of RISs, known as simultaneously transmitting and reflecting RISs (STAR-RISs) have been proposed, where the incident signals can be simultaneously transmitted and reflected towards users roaming at both sides of RISs \cite{liu2021star}. Therefore, compared to reflecting-only RISs, STAR-RISs are capable of achieving full-space coverage as well as introducing new degrees-of-freedom (DoFs) for enhancing the performance \cite{liu2021simultaneously}.
	
	On the other hand, the power-domain (PD) non-orthogonal multiple access (NOMA) has been widely advocated for enhancing the spectral efficiency (SE) and connectivity \cite{jia2019power,liu2017non,vaezi2019interplay,yuan2021noma}. By serving multiple users in the same time/frequency/code resource
	block, NOMA improves an efficient use of the spectrum efficiency beyond that of conventional orthogonal multiple access (OMA) techniques \cite{yuan2020iterative,wang2018beamforming,singhanalysis}. Therefore, the application of NOMA in STAR-RIS assisted networks has been envisioned as a promising network structure, where substantial benefits can be achieved: NOMA  makes efficient use of the spectrum in STAR-RISs aided networks; STAR-RISs are beneficial to NOMA for offering full-space coverage, while improving diversity gain and the decoding order flexibility of NOMA \cite{tang2020physical}.
	\vspace{-15pt}
	\subsection{State-of-the-art}
	\subsubsection{Studies on STAR-RISs}
	Recently, STAR-RISs and their diverse variants have emerged as promising techniques for networking performance improvement \cite{liu2021reconfigurable, pan2021reconfigurable,mu2021simultaneously,xu2021star}.
	In \cite{mu2021simultaneously}, Mu \emph{et al.} studied the power consumption minimization problem of STAR-RIS aided unicast and multicast transmission systems. In \cite{xu2021star}, Xu \emph{et al.} proposed a pair of channel models for the near- and far-field regions of STAR-RIS assisted networks and the corresponding outage probability expressions were derived. In \cite{9525400}, Niu \emph{et al.\
	} proposed a joint passive and active beamforming (BF) design in the downlink of STAR-RIS assisted networks for the maximization of the weighted sum secrecy rate. 
	
	\subsubsection {NOMA in RIS/STAR-RIS-assisted Networks}
	The performance gains of combining NOMA and RIS/STAR-RIS have been intensively investigated \cite{liu2020reconfigurable,zhang2020downlink,9472958,zhang2020robust,ni2021star,Houajoint2021}. In \cite{liu2020reconfigurable}, Liu \emph{et al.} investigated the advantages of employing both beamformer-based and cluster-based NOMA strategies in RIS-aided multi-user networks, where both distributed and centralized RIS deployment were considered. In \cite{zhang2020downlink}, Zhang \emph{et al.} derived the closed-form coverage probability expressions of RIS-aided NOMA networks, quantifying the performance enhancement of employing RISs. In \cite{9472958}, Xiu \emph{et al.} studied the achievable sum-rate maximization problem of RIS-aided mmWave NOMA systems by jointly optimizing the power allocation, phase shifts, and BF. In \cite{zhang2020robust}, Zhang \emph{et al.} considered the energy efficiency maximization problem of the MISO RIS-NOMA downlink by alternately optimizing the BF at both the BS and the RIS. As a brand-new topic, the combination of STAR-RISs and NOMA has the potential of outperforming RIS-NOMA networks in terms of SE \cite{ni2021star}. In \cite{Houajoint2021}, Hou \emph{et al.} proposed a STAR-RIS-aided coordinated multi-point transmission (CoMP) assisted NOMA system, where the active BF, passive BF, and detection vectors are jointly designed for signal power enhancement and interference cancellation. Ni \emph{et al.}  \cite{ni2021star} investigated the STAR-RIS-aided uplink of heterogeneous networks employing NOMA schemes, where a new successive signal processing design was proposed for uplink interference cancellation. 
	In \cite{zuo2021joint}, Zuo \emph{et al.} alternately optimized the power allocation, the active and passive BF at the BS and the STAR-RIS in the downlink of STAR-RIS-NOMA systems for sum-rate maximization.
	\vspace{-20pt} 
	\subsection{Motivations and Contributions}
	Although some authors have focused their attention on the optimization of STAR-RIS assisted NOMA systems, the location deployment problem of the STAR-RIS has not been considered. In STAR-RIS assisted networks, the received signal suffers from the propagation loss of both the BS-STAR-RIS link and of the STAR-RIS-user link, often referred to in jargon as the "double fading" \cite{mu2021joint}, which is closely related to the location of the STAR-RIS. Additionally, in contrast to the reflection-only RIS whose total power is reflected regardless of its location, the specific positions of STAR-RIS is the 
	main factor influciencing its transmission/reflection matrix design (i.e., the ratio of power used for transmission/reflection). Hence it is essential to optimize the location of the STAR-RIS for getting the most out of its advantages.

	Therefore, we focus our attention on the location design of the STAR-RIS and also take into account the joint active and passive BF optimization for maximizing the weighted sum-rate (WSR) of STAR-RIS assisted NOMA systems. The main contributions of this paper are detailed as follows:
	\begin{itemize}
		\item  We investigate a STAR-RIS assisted NOMA communication system, in which the STAR-RIS is deployed for assisting the communication between the BS and users. Depending
		on whether a beamformer serves a single or multiple users, the beamformer-based and cluster-based NOMA strategies are considered, respectively. Accordingly, we formulate the joint beamforming and position design for our WSR maximization problem of both strategies, subjected to the specific SIC decoding order, transmission/reflection power constraints, and minimal required rates of users.
		\item  For the beamformer-based NOMA, we decompose the NP-hard WSR maximization problem formulated into several sub-problems and develop an alternating optimization (AO) based algorithm for jointly optimizing the deployment location of the STAR-RIS, the active BF at the BS, and the passive BF at the STAR-RIS. Specifically, the successive convex approximation (SCA) and semi-definite programming (SDP) methods are employed for addressing these sub-problems, and a two-step iterative algorithm is designed for determining the STAR-RIS location. 
		
		\item For the cluster-based NOMA strategy, the users in a cluster are served by a common beamformer and they are distinguished by their different power allocation factors (PAFs), which have to be optimized in addition to the active BF, the passive BF, and the deployment location. Then our WSR maximization problem is formulated, an AO based algorithm is proposed for solving the resultant non-convex problem, while taking into account the decoding order design.

		\item Our numerical results demonstrate that 1) our proposed STAR-RIS assisted NOMA system outperforms both its OMA and reflection-only RIS counterparts; 2) the deployment optimization of STAR-RIS enhances the WSR; 3) 
		the STAR-RIS deployment location strategy varies from different multiple access schemes. To be specific, OMA prefers symmetric while NOMA prefers asymmetric deployment\footnote{deployed near a specific user.} among users.
		
	\end{itemize}
	%
	\vspace{-20pt}
	\subsection{Organization and Notation}
	The rest of this paper is organized as follows. The system model and problem formulations are introduced in Section II. The joint optimization design of both beamformer-based and cluster-based NOMA is detailed in Section \RNum{3} and Section \RNum{4}, respectively. In Section \RNum{5}, our numerical results are discussed, demonstrating the performance enhancements attained. Finally, Section \RNum{6} concludes the paper.
	
	\emph{Notation:} Scalars are denoted by lower-case letters. Vectors and matrices are denoted by bold-face lower-case and upper-case letters. $ \left(\cdot\right)^{T} $, $ \left(\cdot\right)^{\dag} $, Tr$ \left(\cdot\right) $ and Rank$ \left(\cdot\right) $ stand for the transpose, Hermitian transpose, trace and rank of a
	matrix, $ \left\|  \cdot  \right\| $ and $ \left|  \cdot  \right| $ denote the Euclidean norm and absolute value, $ \mathbb{C}^{M \times N} $ defines the complex
	space of $ M \times N $, while $ \mathbf{A} \succeq 0 $ means that matrix $ \mathbf{A} $ is a positive semi-definite matrix.
	\vspace{-13pt}
	\section{System model and Problem Formulation}
\begin{figure}[ht]
		\centering
		\includegraphics[width=0.45\linewidth]{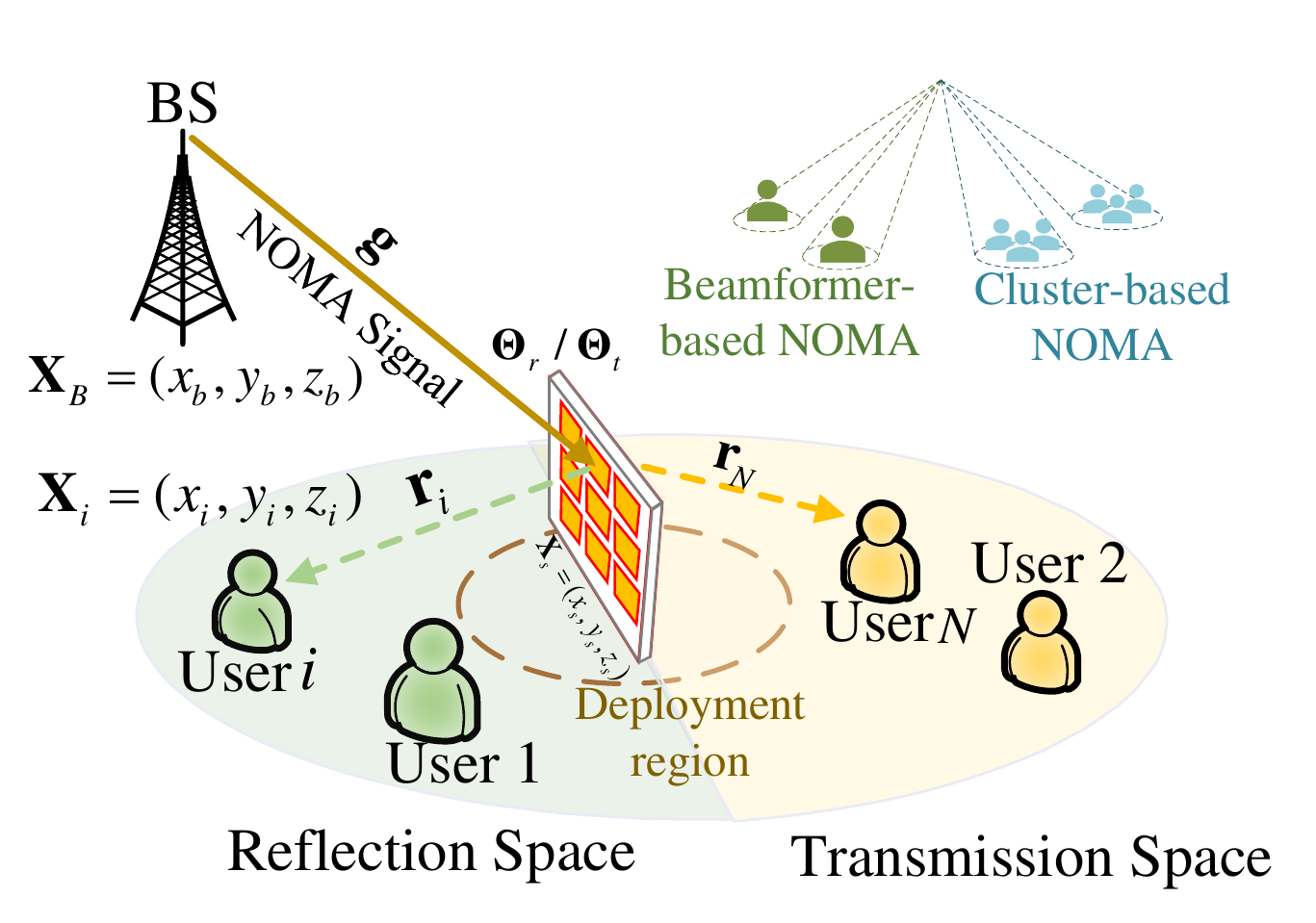}
		\caption{The STAR-RIS assisted NOMA system.}
		\label{system1}
	\end{figure}

	We consider the STAR-RIS assisted NOMA downlink illustrated in Fig. \ref{system1}, where $N$ single-antenna users are served by an $ N_t $-antenna BS. Specifically, as shown in Fig. \ref{system1}, each user is served with a BF in beamformer-based NOMA, while in cluster-based NOMA, $N$ users are assigned into $ N_c $ clusters and the users in each cluster are served by a common BF. Since we focus our attention on the location of the STAR-RIS, we define the location of the BS at $ {\mathbf{X}}_B = (x_b, y_b, z_b) $, the location of the STAR-RIS at $ {\mathbf{s}} = (x_s, y_s, z_s) $, and the location of the $ n $-th user at $ {\mathbf{X}}_{n} = (x_{n}, y_{n}, z_{n}) $. The predefined deployment region of the STAR-RIS is defined as $x_{\min} \le x_s \le x_{\max}$, $y_{\min} \le y_s \le y_{\max}$ and $z_{\min} \le z_s \le z_{\max}$.  Furthermore, we assume an energy splitting (ES) STAR-RIS composed of $M = M_vM_h$ STAR elements, whose transmission and reflection matrices are defined as $ \mathbf{\Theta}_{t}=\operatorname{diag}\left(\mathbf{v}_t\right) $ and $\mathbf{\Theta}_{r}=\operatorname{diag}\left(\mathbf{v}_r\right) $, where  $ \mathbf{v}_{t}=[\sqrt{\beta_{1}^{t}} e^{j \theta_{1}^{t}}, \sqrt{\beta_{2}^{t}} e^{j \theta_{2}^{t}}, \ldots, \sqrt{\beta_{M}^{t}} e^{j \theta_{M}^{t}}] $, $\mathbf{v}_{r}=[\sqrt{\beta_{1}^{r}} e^{j \theta_{1}^{r}}, \sqrt{\beta_{2}^{r}} e^{j \theta_{2}^{r}}, \ldots, \sqrt{\beta_{M}^{r}} e^{j \theta_{M}^{r}}]$, $ \{\sqrt{\beta^t_m}, \sqrt{\beta^r_m} \}$ and $ \{\theta_m^t, \theta_m^r\} $ denote the transmission/reflection amplitudes and phase shift adjustments of the $ m $-th element, which satisfy $ \beta_m^t, \beta_m^r \in [0,1]$, $ \beta_m^t+\beta_m^r=1 $, and $\theta_{m}^{t}, \theta_{m}^{r} \in[0,2 \pi), \forall m  \in  {\cal{M}} \buildrel \Delta \over =  \{1,2,...,M\}$.
	
	The channels of the BS-STAR-RIS link and the STAR-RIS-user link are modeled as Rician fading channels, which can be expressed as \cite{mu2021joint}
	\begin{equation}
		\begin{aligned}
			&\mathbf{G}=\sqrt{\frac{\beta}{1+\beta}} \mathbf{g}^{\operatorname{LoS}}+\sqrt{\frac{1}{1+\beta}} \mathbf{g}^{\mathrm{NLoS}} , \\
			&\mathbf{r}_{n}=\sqrt{\frac{\beta}{1+\beta}} \mathbf{r}_{n}^{\mathrm{LoS}}+\sqrt{\frac{1}{1+\beta}} \mathbf{r}_{n}^{\mathrm{NLoS}} ,
		\end{aligned}
	\end{equation}
	where $ \mathbf{G} \in \mathbb{C}^M$ is the channel between the BS and the STAR-RIS, $ \mathbf{r}_{n} $ denotes the channel between the STAR-RIS and the $n$-th user, $ \mathbf{g}^{\rm{LoS}} $ and $ \mathbf{r}_{n}^{\rm{LoS}} $ are the deterministic LoS components, $ \mathbf{g}^{\rm{NLoS}} $ and $ \mathbf{r}_{n}^{\rm{NLoS}} $ denote the Rayleigh distributed non-line-of-sight (NLoS) components \cite{mu2021joint}, while $\beta$ is the Rician factor.
	Furthermore, the large-scale path-loss between the BS the $n$-th user is expressed as
	$ 	L_{n}=d_{B-S}^{-\alpha}d_{S-{U_{n}}}^{-\alpha}, $
	where $ \alpha $ is the path-loss component, $ d_{B-S} $ and $ d_{S-{U_{n}}} $ denote the distance between the BS and the STAR-RIS and between the STAR-RIS and the $ n $-th user.
	\vspace{-20pt}
	{\subsection{Problem Formulation of Beamformer-based NOMA}}
	In beamformer-based STAR-RIS-NOMA, the BS serves each user by a beamformer and the STAR-RIS is employed to transmit/reflect the BS's downlink signals to users. The $ i $-th user's received signal can be expressed as
	\begin{equation}
		{y_i} = \sum\limits_{i = 1}^N\left( {\sqrt {{L_i}} {\bf{r}}_i^\dag{{\bf{\Theta }}_{i,p}}{\bf{G}}} \right) {{{\bf{w}}_i}{s_i}}  + {n_i},
	\end{equation}
	where $ \mathbf{\Theta}_{i,p} $, $p \in \{t,r\}$ denotes the transmission/reflection matrix between the STAR-RIS and the $ i $-th user, we have $ \mathbf{\Theta}_{i,p} =  \mathbf{\Theta}_{t}$ when the $ i $-th user is located in the transmission area, and $ \mathbf{\Theta}_{i,p} =  \mathbf{\Theta}_{r}$ when the $  i $-th user is in the reflection area, $ s_i $ denotes the $ i $-th user's signal, $ n_i \sim {\cal{CN}}\left(0,\sigma^2\right)$ is the additive white Gaussian noise (AWGN), and $ \mathbf{w}_i $ is the active BF vector of the $ i $-th user. According to the NOMA principle, successive interference cancellation (SIC) is employed by the users to remove all other users' interference in order. We define the decoding order of the $i$-th user by $\pi(i)$, and $\pi(k) > \pi(i)$ indicates the user $k$ will first detect the signal of user $i$ before detecting its own signals. Let ${\cal{J}}_i \buildrel \Delta \over =  \{k|\pi(k) > \pi(i)\}$ denotes the set of users detected after the $ i $-th user. Then the achievable rate of the $ i $-th user in decoding its own signal is given by
	\begin{equation}
		{{R_{i \to i} = {\log _2}\left( {1 + \frac{{{L_i}{{\left| {{\bf{r}}_i^\dag{{\bf{\Theta }}_{i,p}}{\bf{G}}{{\bf{w}}_i}} \right|}^2}}}{{\sum\limits_{j \in {\cal{J}}_i} {{L_i}{{\left| {{\bf{r}}_i^\dag{{\bf{\Theta }}_{i,p}}{\bf{G}}{{\bf{w}}_j}} \right|}^2}}  + \sigma _i^2}}} \right),}}
		\label{pp04}
	\end{equation}
	and the rate for the $k$-th user ($k \in {\cal{J}}_i$) in detecting the $ i $-th user's signal can be expressed as
	\begin{equation}
		{R_{k \to i} = {\log _2}\left( {1 + \frac{{{L_k}{{\left| {{\bf{r}}_k^\dag{{\bf{\Theta }}_{k,p}}{\bf{G}}{{\bf{w}}_i}} \right|}^2}}}{{\sum\limits_{j \in {\cal{J}}_i} {{L_k}{{\left| {{\bf{r}}_k^\dag{{\bf{\Theta }}_{k,p}}{\bf{G}}{{\bf{w}}_j}} \right|}^2}}  + \sigma _k^2}}} \right).}
	\end{equation}
	The WSR maximization problem can be formulated by jointly designing the active BF at the BS, the passive BF and the location of the STAR-RIS as
	\begin{subequations} \label{p5}
		\begin{align}
			\mathop {\max }\limits_{{{\bf{w}}_i}, \mathbf{v}_p, \mathbf{s}}~ &\sum\limits_{i=1}^{N}{\eta _i}{R_{i \to i}}
			\label{1-1}\\
			{\text{s.t.}}\quad
			& R_{i \to i} \ge R_{\min}, ~\forall i \in \cal{I},
			\label{1-1a}\\
			& {R_{k \to i} \ge R_{i \to i},} ~\forall i \in {\cal{I}}, k \in {\cal{J}}_i,
			\label{1-2}\\
			&  {{L_i\left| {{\bf{r}}_i^\dag{{\bf{\Theta }}_{i,p}}{\bf{G}}\mathbf{w}_i} \right|^2} \le L_j {\left| {{\bf{r}}_j^\dag{{\bf{\Theta }}_{j,p}}{\bf{G}}\mathbf{w}_j} \right|^2},~ \forall i \in {\cal{I}}, j \in {\cal{J}}_i,}
			\label{1-3}\\
			& \sum\limits_{i = 1}^N {{{\left\| {{{\bf{w}}_i}} \right\|}^2}}  \le {P_{\max }},
			\label{1-4}\\
			& \mathbf{s}\left(x,y\right) \in {\cal{X}} \times {\cal{Y}},
			\label{1-5}\\
			& \beta^p_m \in [0,1], \theta_m^p \in \left[0, 2\pi \right), ~\forall m \in {\cal{M}}, p\in \{t,r\},
			\label{1-6}\\
			& \beta^t_m +\beta ^r_m =1, ~\forall m \in {\cal{M}},
			\label{1-7}
		\end{align}
	\end{subequations}
	where $ {\cal{I}} = \{1,2,...,N\} $, $\mathbf{s}\left(x,y\right)$ defines the x- and y-axis location of the STAR-RIS, while $ \eta_i $ denotes the rate weight of the $ i $-th user. Constraint (\ref{1-1a}) is the minimal rate requirement of the $i $-th user, (\ref{1-2})
	guarantees that the SIC can be successfully performed, (\ref{1-3}) guarantees the fairness among users, (\ref{1-4}) is the total transmit power constraint of the BS, (\ref{1-5}) is the STAR-RIS deployment constraint, while (\ref{1-6}) and (\ref{1-7}) are the transmission and reflection power constraints of the STAR-RIS. 
	\vspace{-3pt}
	{\subsection{Problem Formulation of Cluster-based NOMA}}
	\vspace{-5pt}
	In this part, the signal model and the WSR maximization problem of the cluster-based NOMA are formulated. The received signal of the $ i $-th user in the $n$-th cluster can be expressed as
	
	\begin{equation} \label{eq7}
		{y_{n,i}} = \sum\limits_{n=1}^{N_c}\left( {\sqrt {{L_{n,i}}} {\bf{r}}_{n,i}^\dag{{\bf{\Theta }}_{n,p}}{\bf{G}}\mathbf{w}_n} \right)\sum\limits_{i = 1}^{K} {\sqrt {{p_{n,i}}} {s_{n,i}}}  + {n_{n,i}},
	\end{equation}
	where $N$ users are assigned into $N_c$ clusters with $K$ users in a cluster, $L_{n,i}$ denotes the propagation path-loss of the $i$-th user in the $n$-th cluster.  Furthermore, ${\bf{r}}_{n,i}$ is the channel between the STAR-RIS and the $i$-th user in the $n$-th cluster, ${\bf{\Theta }}_{n,p}$, $ p \in \{t,r\} $ is the transmission and reflection matrix, which can be expressed as ${\bf{\Theta }}_{n,p} = {\bf{\Theta }}_{t}$ when the $n$-th cluster is located in the transmission area, and ${\bf{\Theta}}_{n,p} = {\bf{\Theta }}_{r}$ when the $n$-th cluster is in the reflection area, still {\color{red}regaining} to \eqref{eq7}, $\mathbf{w}_n$ is the BF vector of the $n$-th cluster, $p_{n,i}$ and $s_{n,i}$ are the PAF and the desired signal of the $i$-th user in the $n$-th cluster, and $n_{n,i} \sim {\cal{CN}}\left(0,\sigma^2\right)$ denotes the AWGN. For any two users $k$ and $i$ in the cluster $n$, $\pi(n,k) >\pi(n,i)$ indicates that user $i$ is detected before user $k$. Then the achievable rate of the $i$-th user in the $n$-th cluster is given by
	\begin{equation}
		{{\tilde R_{n,{i\to i}}} = {\log _2}\left( {1 + \frac{{{p_{n,i}}{L_{n,i}}|\mathbf{r}_{n,i}^\dag\boldsymbol{\Theta}_{n,p}\mathbf{G}\mathbf{w}_n|^2}}{{\sum\limits_{j \in {\cal{J}}_{n,i}} {{p_{n,j}}{L_{n,i}}|\mathbf{r}_{n,i}^\dag\boldsymbol{\Theta}_{n,p}\mathbf{G}\mathbf{w}_n|^2}  }+L_{n,i}\sum\limits_{n'\ne n}|\mathbf{r}_{n,i}^\dag\boldsymbol{\Theta}_{n,p}\mathbf{G}\mathbf{w}_n'|^2+ \sigma _{n,i}^2}} \right),}
	\end{equation}
	where ${\cal{J}}_{n,i}$ denotes the set consists of users detected after the $i$-th user in the $n$-th cluster, for all $k \in {\cal{J}}_{n,i}$. Then the rate for the $k$-th user to detect the $i$-th user's signal can be expressed as
	\begin{equation}
		{{\tilde R_{n,{k\to i}}} = {\log _2}\left( {1 + \frac{{{p_{n,i}}{L_{n,k}}|\mathbf{r}_{n,k}^\dag\boldsymbol{\Theta}_{n,p}\mathbf{G}\mathbf{w}_n|^2}}{{\sum\limits_{j \in {\cal{J}}_i} {{p_{n,j}}{L_{n,k}}|\mathbf{r}_{n,i}^\dag\boldsymbol{\Theta}_{n,p}\mathbf{G}\mathbf{w}_n|^2}  }+L_{n,k}\sum\limits_{n'\ne n}|\mathbf{r}_{n,k}^\dag\boldsymbol{\Theta}_{n,p}\mathbf{G}\mathbf{w}_n'|^2+ \sigma _{n,k}^2}} \right).}
	\end{equation}
	
	The WSR maximization problem can be formulated as
	\begin{subequations}  	\label{p40}
		\begin{align}
			\mathop {\max }\limits_{ {p_{n,i}}, {{\bf{V}}_p},\hfill\atop{\bf{s}},{\bf{W}}_n\hfill} ~& \sum\limits_{n=1}^{ N_c}\sum\limits_{i = 1}^K {{\eta _{n,i}}{\tilde R_{n,{i \to i}}}}
			\label{7-1}\\
			{\text{s.t.}}\;\;
			& \mathbf{s}\left(x,y\right) \in \cal{X}\times\cal{Y},
			\label{7-2}\\
			& \tilde R_{n,{i\to i}} \ge R^{\min}_{n,{i \to i}},
			\label{7-3}\\
			& \tilde R_{n,{k\to i}} \ge \tilde R_{n,{i \to i}},
			\label{7-4}\\
			& \sum\limits_{i = 1}^{K} {{p_{n,i}}}  = 1,
			\label{7-6}\\
			& \sum\limits_{i = 1}^{K} \|\mathbf{w}_n\|^2 \le P_{\max}, 
			\label{7-7}\\
			& \beta_m^p \in [0,1], \theta_m^p \in \left[ {0,2\pi } \right),~ m \in {\cal{M}}, p \in \{t,r\},
			\label{7-8}\\
			& \beta_m^t + \beta_m^r = 1,~ m \in {\cal{M}},
			\label{7-9}
		\end{align}
	\end{subequations}
	where $\eta _{n,i}$ is the weight of the $i$-th user in the $n$-th cluster, (\ref{7-2}) is the deployment constraint, (\ref{7-3}) is the users' minimal rate constraint, (\ref{7-4}) is to guarantee successful SIC detection, (\ref{7-6}) is the power allocation constraint in a cluster, (\ref{7-7}) is the total transmit power constraint at the BS, while (\ref{7-8}) and (\ref{7-9}) are constraints of the passive BF at the STAR-RIS.
	\section{Proposed Optimization Algorithm for Beamformer-based NOMA}
	\label{section3}
	In this section, we proposed an AO based algorithm to solve the WSR maximization problem (\ref{p5}), where the active BF at the BS, the passive BF, and the location of the STAR-RIS are alternately optimized with all the other variables fixed. Exhaustive search can be employed to obtain the optimal decoding order, but we assume a given detect order for simplicity, and decompose the original problem (\ref{p5}) into several sub-problems as follows.
	\vspace{-20pt}
	\subsection{The Sub-problem of Active BF Design}
	\label{subsection1}
	The problem of optimizing the active BF at the BS with the fixed STAR-RIS location $\mathbf{s}$ and the passive BF $ \mathbf{v}_p $, $p \in \{t,r\}$ is formulated as
	\begin{subequations}
		\begin{align}
			\mathop {\max }\limits_{{{\bf{w}}_i}}~&\sum\limits^{N}_{i=1} {\eta _i}{R_{i\to i}}
			\label{p2-1}\\
			{\text{s.t.}}\quad
			& L_i {{| {{\bf{r}}_i^\dag{{\bf{\Theta }}_{i,p}}{\bf{G}}{{\bf{w}}_i}} |^2} \le L_j {\left| {{\bf{r}}_j^\dag{{\bf{\Theta }}_{j,p}}{\bf{G}}{{\bf{w}}_j}} \right|^2}, ~\forall i \in {\cal{I}}, j \in {\cal{J}}_i,}
			\label{2-2}\\
			& {R_{k \to i} \ge R_{i \to i},} ~\forall i \in {\cal{I}}, k \in {\cal{J}}_i,
			\label{2-3}\\
			& {R_{i \to i} \ge R_{\min},} ~\forall i \in {\cal{I}},
			\label{2-3a}\\
			& \sum\limits_{i = 1}^K {{{\left\| {{{\bf{w}}_i}} \right\|}^2}}  \le {P_{\max }}.
			\label{2-4}
		\end{align}
	\end{subequations}
	Then the optimization problem can be formulated by employing the semi-definite programming (SDP) method as
	\begin{subequations} \label{3}
		\begin{align}
			\mathop {\max }\limits_{{{{\bf{W}}_i}}}~&\sum\limits_{i = 1}^N {\eta _i} R_{i \to i}
			\label{p3-1}\\
			{\text{s.t.}}\quad
			&  {{\rm{Tr}}\left( {{{\bf{W}}_i}{\bf{H}}_i^b} \right) \le {\rm{Tr}}\left( {{{\bf{W}}_j}{\bf{H}}_j^b} \right), ~ \forall i \in {\cal{I}}, j \in {\cal{J}}_i,}
			\label{p3-2}\\
			& \sum\limits_{i = 1}^K {{\rm{Tr}}\left( {{{\bf{W}}_i}} \right)}  \le {P_{\max }},
			\label{p3-4}\\
			& {\rm{Rank}} \left( {\mathbf{W}}_i\right) = 1, ~\forall i \in \cal{I},
			\label{p3-5}\\
			& \mathbf{W}_i \succeq 0,~\forall i \in \cal{I},\label{p3-6}\\
			& {\text{(\ref{2-3}), (\ref{2-3a})}},
		\end{align}
	\end{subequations}
	where $ \mathbf{h}^b_i = \sqrt{L_i}\mathbf{r}_i\mathbf{\Theta}_{i,p}\mathbf{G}$, $ \mathbf{H}^b_i = (\mathbf{h}^b_i)^{\dag}\mathbf{h}^b_i $, $R_{i \to i} = \frac{{\rm{Tr}}\left( {{{\bf{W}}_i}{\bf{H}}_i^b} \right)}{\sum\limits_{j \in {\cal{J}}_i }{{\rm{Tr}}\left( {{{\bf{W}}_j}{\bf{H}}_i^b} \right) + \sigma _i^2}}$. To address the non-convex problem (\ref{3}), we introduce the slack variables $\{x_i\}$ and $\{y_i\}$ so that
	\begin{equation}
		\frac{1}{{{x_{i,i}}}} = {\rm{Tr}}\left( {{{\bf{W}}_i}{\bf{H}}_i^b} \right),~\forall i \in \cal{I},
		\label{133}
	\end{equation}
	\begin{equation}
		{y_{i,i}} = \sum\limits_{j \in {\cal{J}}_i }{{\rm{Tr}}\left( {{{\bf{W}}_j}{\bf{H}}_i^b} \right) + \sigma _i^2},~\forall i \in \cal{I}.
		\label{134}
	\end{equation}
	Then the resultant optimization problem can be formulated as
	\begin{subequations} \label{p4}
		\begin{align}
			\mathop {\max }\limits_{{{{\bf{W}}_i}}, R_{i \to  i}, x_{i,i}, y_{i,i}}~&\sum\limits_{i = 1}^N {\eta _i} R_{i \to i}
			\label{r4-1}\\
			{\text{s.t.}}\quad
			& R_{i \to i} \le \log_2\left(1+\frac{1}{x_{i,i}y_{i,i}}\right), ~\forall i \in \cal{I},\label{r4-2}\\
			&  \frac{1}{x_{i,i}} \le {\rm{Tr}}\left( {{{\bf{W}}_i}{\bf{H}}_i^b} \right),~\forall i \in \cal{I},\label{r4-3}\\
			& y_{i,i} \ge  \sum\limits_{j \in {\cal{J}}_i }{{\rm{Tr}}\left( {{{\bf{W}}_j}{\bf{H}}_i^b} \right) + \sigma _i^2},~\forall i \in \cal{I},\label{r4-4}\\
			& R_{i \to i} \le \log_2\left(1+\frac{1}{x_{i,k}y_{i,k}}\right), ~\forall i \in {\cal{I}}, k \in {\cal{J}}_i,\label{r4-5}\\
			&  R_{i \to i} \ge R_{\min},~\forall i \in {\cal{I}},\label{r4-6}\\
			& \text{(\ref{p3-2})-(\ref{p3-5})},
		\end{align}
	\end{subequations}
	where $	\frac{1}{{{x_{i,k}}}} = {\rm{Tr}}\left( {{{\bf{W}}_i}{\bf{H}}_k^b} \right)$, ${y_{i,k}} = \sum\limits_{j \in {\cal{J}}_i }{{\rm{Tr}}\left( {{{\bf{W}}_j}{\bf{H}}_k^b} \right) + \sigma _i^2}$, $\forall i \in \cal{I}$, $k \in {\cal{J}}_i$.
	However, problem (\ref{p4}) is non-convex due to the non-convexity of the constraints (\ref{p3-5}), (\ref{r4-2}), and (\ref{r4-5}).
	As proved in \cite{mu2020exploiting}, $ \log_2\left(1+\frac{1}{xy}\right) $ is a convex function with respect to $x$ and $y$ when $x>0$, $y > 0$, hence, a linearized lower bound of the right-hand-side of (\ref{r4-2}) and (\ref{r4-5}) can be obtained by employing the first-order Taylor expansion at the feasible point $(\tilde x_i, \tilde y_i)$ as
	\begin{equation}
		\log_2\left(1+\frac{1}{x_{i,i}y_{i,i}}\right) \ge \tilde R_{i \to i}  \buildrel \Delta \over =  {\log _2}\left( {1 + \frac{1}{{{{\tilde x}_{i,i}}{{\tilde y}_{i,i}}}}} \right) - \frac{{{{\log }_2}e\left( {{x_{i,i}} - {{\tilde x}_{i,i}}} \right)}}{{{{\tilde x}_{i,i}}^2{{\tilde y}_{i,i}} + {{\tilde x}_{i,i}}}} - \frac{{{{\log }_2}e\left( {{y_{i,i}} - {{\tilde y}_{i,i}}} \right)}}{{{{\tilde x}_{i,i}}{{\tilde y}_{i,i}}^2 + {{\tilde y}_{i,i}}}},~ \forall i \in  \cal{I}.
		\label{p11}
	\end{equation}
	
	Therefore, the optimization problem can be reformulated as
	
	\begin{subequations} \label{p6}
		\begin{align}
			\mathop {\max }\limits_{{{{\bf{W}}_i}}, R_{i \to  i}, x_{i,i}, y_{i,i}}~&\sum\limits_{i = 1}^N {\eta _i} R_{i \to i}
			\label{r5-1}\\
			{\text{s.t.}}\quad
			& R_{i \to i} \le \tilde R_{i \to i}, ~\forall i \in \cal{I}, \label{r5-2} \\
			& R_{i \to i} \le  \tilde R_{k \to i}, ~\forall i \in {\cal{I}}, k \in {\cal{J}}_i,\label{r5-3}\\
			&  R_{i \to i} \ge R_{\min},~\forall i \in {\cal{I}},\label{r5-4}\\
			& \text{(\ref{p3-2})-(\ref{p3-6}), (\ref{r4-3}), (\ref{r4-4}).}
		\end{align}
	\end{subequations}
	
	Note that the rank-one constraint (\ref{p3-5}) is provably satisfied, hence we drop it in order to obtain a relaxed version, which is tractable and can be solved by the CVX.
	\begin{remark}
		The optimal solution $ {\mathbf{W}_i}^* $ of problem (\ref{p6}) obtained without the rank-one constraint always satisfies the rank-one constraint. The proof can be found in \cite{zuo2021joint}.
	\end{remark}
	According to (\ref{p6}), our algorithm proposed for obtaining the active BF $\mathbf{w}_i$ is summarized in {\bf{Algorithm}} 1, where $\kappa$ denotes the iteration index of the AO procedure, in which the solution of the $(\kappa-1)$-st iteration is used as the input of the $\kappa$-th iteration, $\kappa_b$ is the iteration index of {\bf{Algorithm}} 1 for obtaining a locally optimal solution.
	\vspace{-10pt}
	\begin{algorithm}
		\LinesNumbered
		\caption{Proposed Algorithm for Optimizing the Active BF in the $\kappa$-th AO Iteration.}
		\KwIn {	The passive BF $ \mathbf{v}_t^{(\kappa-1)} $ and $ \mathbf{v}_r^{(\kappa-1)} $, the optimized location of STAR-RIS $\mathbf{s}^{(\kappa-1)}$.}
		Set the tolerance of iteration accuracy $ \tilde \delta_1 $, max iteration time $ I^{\max}_1 $, initial iteration number $ \kappa_b=0 $, and initial feasible points $\{\tilde x_{i,i}\}, \{\tilde y_{i,i}\}$\;
		\While{$\kappa_b \le I^{\max}_1$ and $ \delta > \tilde \delta_1 $}{
			$ \kappa_b = \kappa_b+1 $\;
			Solve (\ref{p6}) to obtain the solution $\mathbf{W}_i^{(\kappa_b)}$\;
			Update $\{\tilde x_{i,i}\}, \{\tilde y_{i,i}\}$, and objective value $ F^{(\kappa_b)} $, $ \delta = \left|F_{tr}^{(\kappa_b)}-F_{tr}^{(\kappa_b-1)}\right| $.
		}
		\KwOut {$\mathbf{W}_i^{(\kappa)}$.}
	\end{algorithm}
	\vspace{-35pt}
	\subsection{The Sub-problem of Passive BF Design}
	In this part, we optimize the passive BF $ \mathbf{v}_p ,~p\in \{{t,r}\}$ at the STAR-RIS with given $ \mathbf{w}_i $ and location $\mathbf{s}$. Let us define $\mathbf{V}_{i,p} = \mathbf{v}_{i,p}^\dag\mathbf{v}_{i,p}$, which satisfies $\mathbf{V}_{i,p} \succeq 0$ and Rank$(\mathbf{V}_{i,p})=1$, where $\mathbf{V}_{i,p}  = \mathbf{V}_t$ when the $i$-th user is located in the transmission area, and $\mathbf{V}_{i,p}  = \mathbf{V}_r$ when the $i$-th user is deployed in the reflection area. Hence, $R_{i \to i}$ can be formulated as
	\begin{equation}
		R_{i \to i} = {{\log _2}\left( {1 + \frac{{{\rm{Tr}}\left( {{{\bf{V}}_{i,p}}{\bf{H}}_{i,i}^v} \right)}}{{\sum\limits_{ j \in {\cal {J}}_i} {{\rm{Tr}}\left( {{{\bf{V}}_{i,p}}{\bf{H}}_{i,j}^v} \right)}  + \sigma _i^2}}} \right)},
	\end{equation}
	where $\mathbf{h}_{i,i}^v = \sqrt{L_i}{\rm{diag}}(\mathbf{r}_i^{\dag})\mathbf{G}\mathbf{w}_i$, $\mathbf{h}_{i,j}^v = \sqrt{L_i}{\rm{diag}}(\mathbf{r}_i^{\dag})\mathbf{G}\mathbf{w}_j$, $\mathbf{H}_{i,i}^{\dag} = \mathbf{h}_{i,i}^v(\mathbf{h}_{i,i}^v)^\dag$, and $\mathbf{H}_{i,j}^v = \mathbf{h}_{i,j}^v(\mathbf{h}_{i,j}^v)^\dag$. 
	Furthermore, according to (\ref{1-6}), we have ${\left[ {{{\bf{V}}_p}} \right]_{m,m}} = \beta _m^p$, where $\beta _m^p$ satisfies $ \beta _m^t + \beta _m^r = 1$ and $0<\beta _m^p<1$, $p \in \{t,r\}$. The corresponding optimization problem can be formulated as
	\begin{subequations} \label{p19}
		\begin{align}
			\mathop {\max }\limits_{{{\bf{V}}_t},{{\bf{V}}_r}, R_{i \to i}} ~& \sum\limits_{i = 1}^N {{\eta_i}} R_{i \to i}
			\label{pp5-1}\\
			{	\text{s.t.}}\quad
			& R_{i \to i} \ge R_{\min},~\forall i \in \cal{I}, \label{pp5-2}\\
			& R_{k \to i} \ge R_{i \to i},~\forall i \in {\cal{I}}, k \in {\cal{J}}_i, \label{pp5-3}\\
			& {{\rm{Tr}}\left( {{{\bf{V}}_{i,p}}{\bf{H}}_{i,i}^v} \right) \le {{\rm{Tr}}\left( {{{\bf{V}}_{j,p}}{\bf{H}}_{j,j}^v} \right)}, \forall i \in {\cal{I}}, j \in {\cal{J}}_i,}
			\label{p5-2}\\
			& {{\bf{V}}_p} \succeq 0, p \in \{t, r\},
			\label{p5-5}\\
			& {\left[ {{{\bf{V}}_p}} \right]_{m,m}} = \beta _m^p, p \in \{t,r\}, \forall m\in {\cal{M}},
			\label{p5-6}\\
			& 0 < \beta _m^p < 1,  p \in \{t,r\}, \forall m \in {\cal{M}},
			\label{p5-7}\\
			& \beta _m^t + \beta _m^r = 1, \forall m\in {\cal{M}},
			\label{p5-8}\\
			& {\rm{Rank}}\left( {{{\bf{V}}_p}} \right) = 1, p \in \{t, r\}.
			\label{p5-9}
		\end{align}
	\end{subequations}
	Similar to Section \ref{subsection1}, we introduce slack variables $ \{A_{i,i}\} $ and $ \{B_{i,i}\} $ to transform the original problem (\ref{p19}) into
	\begin{subequations}
		\begin{align}
			\mathop {\max }\limits_{\boldsymbol{\Phi}^v} ~&\eta_iR_{i \to i}
			\label{p15-1}\\
			{\text{s.t.}}\quad
			&  \frac{1}{{{A_{i,i}}}} \le {\rm{Tr}}\left( {{{\bf{V}}_{i,p}}{\bf{H}}_{i,i}^v} \right),~ \forall i \in {\cal{I}},
			\label{15-2}\\
			& {B_{i,i}} \ge {\sum\limits_{ j \in {\cal{J}}_i} {{\rm{Tr}}\left( {{{\bf{V}}_{i,p}}{\bf{H}}_{i,j}^v} \right)}  + \sigma _i^2}, ~\forall i \in {\cal{I}},
			\label{pp15-3}\\
			& R_{i \to i} \ge  R_{\min},~\forall i \in \cal{I},\label{pp15-5}\\
			& R_{i \to i} \le \log_2\left(1+\frac{1}{A_{i,i}B_{i,i}}\right),~\forall i \in \cal{I}, 
			\label{pp15-4}\\
			& R_{i \to i} \le \log_2\left(1+\frac{1}{A_{i,k}B_{i,k}}\right),~\forall i \in {\cal{I}}, k \in {\cal{J}}_i,
			\label{pp15-6}\\
			&{\text{(\ref{p5-2})-(\ref{p5-9}).}}
		\end{align}
	\end{subequations}
	To address the non-convex constraints (\ref{pp15-4}) and (\ref{pp15-6}), we give the linearized lower bound of their right-hand-side at the feasible point $ \left(\tilde A_{i,i}, \tilde B_{i,i}\right) $ as
	\begin{equation}
		R_{i \to i}^{LB} \buildrel \Delta \over=  {{\log _2}\left( {1 + \frac{1}{{{{\tilde A}_{i,i}}{\tilde B}_{i,i}}}} \right) - \frac{{{{\log }_2}e\left( {{A_{i,i}} - {\tilde A}_{i,i}} \right)}}{{{{\tilde A}_{i,i}}^2{{\tilde B}_{i,i}} + {\tilde A}_{i,i}}} - \frac{{{{\log }_2}e\left( {{B_{i,i}} - {{\tilde B}_{i,i}}} \right)}}{{{{\tilde A}_{i,i}}{{\tilde B}_{i,i}}^2 + {\tilde B}_{i,i}}}},~\forall i \in \cal{I}.
	\end{equation}
	Furthermore, to relax the rank-one constraint (\ref{p5-9}), the sequential rank-one constraint relaxation is employed as \cite{mu2021joint}
	\begin{equation}
		{\rm{max}}\left( {{\rm{eig}}\left( {{{\bf{V}}_p}} \right)} \right) \ge {\varepsilon }{\rm{Tr}}\left( {{{\bf{V}}_p}} \right), p \in \{t, r\},
	\end{equation}
	where $ \max ({\rm{eig}} \left(\mathbf{V}_p\right) ) $ denotes the largest eigenvalue of $ \mathbf{V}_p $, while $ \varepsilon \in [0, 1] $ is a tightness parameter. Finally, we have
	\begin{subequations} \label{p16}
		\begin{align}
			\mathop {\max }\limits_{\boldsymbol{\Phi}_v}~&\eta_iR_{i \to i},
			\label{pp7-1}\\
			{\text{s.t.}}\quad
			&  \mathbf{u}_{\max}\left(\mathbf{V}_p^{(\kappa_v)}\right)^{\dag}\mathbf{V}_p\mathbf{u}_{\max}\left(\mathbf{V}_p^{(\kappa_v)}\right) \ge \varepsilon^{(\kappa_v)} {\rm{Tr}}\left(\mathbf{V}_p\right), p \in \{t, r\},
			\label{pp7-2}\\
			& R_{i \to i} \le  R_{i\to i}^{LB},~\forall i \in\cal{I}, \label{pp7-3}\\
			&R_{i \to i} \le  R_{k\to i}^{LB},~\forall i \in{\cal{I}}, k \in {\cal{J}}_i, \label{pp7-4}\\
			& {\text{(\ref{p5-2})-(\ref{p5-8}), (\ref{15-2})-(\ref{pp15-5}),}}
		\end{align}
	\end{subequations}
	where $\boldsymbol{\Phi}_v\triangleq \{\mathbf{V}_t, \mathbf{V}_r, \{A_{i,i}\}, \{B_{i,i}\}, R_{i \to i}\}$, $ \mathbf{V}_p^{(\kappa_v)} $ is the obtained solution after the $ \kappa_v $-th iteration, $ \mathbf{u}_{\max}\left(\mathbf{V}_p^{(\kappa_v)}\right) $ denotes the eigenvector of $ \mathbf{V}_p^{(\kappa_v)} $ with largest eigenvalue, and the value of $ \varepsilon^{(\kappa_v)} $ can be updated by \cite{mu2021joint} after each iteration.
	We finally arrive at the convex problem (\ref{p16}), which can be solved by CVX.
	
	\vspace{-15pt}
	\subsection{The Sub-problem of Deployment Location Design}
	In this part, we focus on optimizing the deployment location of the STAR-RIS based on the active and passive BF obtained.
	
	Note that in the deployment location optimization procedure, the angle-of-arrival (AoA)/angle-of-departure (AoD) is related to the location, hence the channel $\mathbf{r}_i$ and $\mathbf{G}$ cannot be regarded as constants during the optimization. Therefore, we develop a two-step iterative algorithm, where the location is roughly optimized by only considering the large-scale fading to obtain an initial optimized location first. Then the location is further optimized in vicinity, in which the AoA/AoD can be regarded as constants.
	Therefore, we define $ u_{i,i}=| {{\bf{r}}_i^\dag{{\bf{\Theta }}_{i,p}}{\bf{G}}{{\bf{w}}_i}} |^2$ and $ u_{i,j}=| {{\bf{r}}_i^\dag{{\bf{\Theta }}_{i,p}}{\bf{G}}{{\bf{w}}_j}} |^2$, and $ R_{i \to i} $ can be reformulated as
	$ R_{i \to i} = {\log _2}( 1 + \frac{u_{i,i}}{\sum\limits_{j \in {\cal{J}}_i}u_{i,j}+\frac{\sigma^2_i}{L_i}}) $. The optimization problem can be formulated as
	\begin{subequations} \label{4}
		\begin{align}
			\mathop {\max }\limits_{\bf{s}} ~&{\sum_{i=1}^{N}} {\eta_i}R_{i \to i}
			\label{4-1}\\
			{\text{s.t.}}\quad
			& R_{i\to i } \ge R_{\min}, ~\forall i \in \cal{I}, \label{4-1a}\\
			&{{L_i}|\mathbf{r}_i^\dag\boldsymbol{\Phi}_{i,p}\mathbf{G}\mathbf{w}_i|^2 \le L_j {|\mathbf{r}_j^\dag\boldsymbol{\Phi}_{j,p}\mathbf{G}\mathbf{w}_j|^2}, ~\forall i \in {\cal{I}}, j \in {\cal{J}}_i,}
			\label{4-2}\\
			&R_{k \to i} \ge R_{i \to i}, ~\forall i \in {\cal{I}}, k \in {\cal{J}}_i,
			\label{4-3}\\
			&\mathbf{s} \left(x,y\right) \in \cal{X}\times \cal{Y}.
			\label{4-5}
		\end{align}
	\end{subequations}
	
	To address the non-convex problem (\ref{4}), we first introduce the slack variable $\tau_i = \frac{1}{L_i}$. Then the linearized lower bound of $  R_{i \to i} $ can be found by employing the first-order Taylor expansion at the local point $\tilde \tau_i$ as
	\begin{equation}
		R_{i \to i}^s \triangleq	{\log _2}\left( {1 + \frac{{{u_{i,i}}}}{{\sum\limits_{j \in {\cal {J}}_i} {{u_{i,j}}}  + \tilde \tau _i\sigma _i^2}}} \right) - \frac{{{u_{i,i}}\sigma _i^2{{\log }_2}e\left({\tau _i}- {\tilde \tau _i} \right)}}{{\left( {{u_{i,i}} + \sum\limits_{j \in {\cal {J}}_i}u_{i,j} +\tilde \tau _i\sigma _i^2} \right)\left( {\sum\limits_{j \in {\cal {J}}_i}u_{i,j} + \tilde \tau _i\sigma _i^2} \right)}}, ~\forall i \in \cal{I}.
	\end{equation}
	Then (\ref{4}) can be reformulated as
	\begin{subequations} \label{5}
		\begin{align}
			\mathop {\max }\limits_{\boldsymbol{\Phi}_s }~&\sum\limits_{i = 1}^K {{\eta _i}}  R_{i \to i}^s
			\label{5-1}\\
			{\text{s.t.}}\quad
			& R_{i \to i}^s \ge R_{\min}, \forall i \in { \cal{I}}, \label{5-2}\\
			& 	{\left\| {{\bf{\tilde s}} - {{\bf{X}}_j}} \right\|^2} + 2\left( {{\bf{\tilde s}} - {{\bf{X}}_j}} \right)\left( {{\bf{s}} - {\bf{\tilde s}}} \right)^T \ge {\left( {\frac{{{u_{i,i}}}}{{{u_{j,j}}}}} \right)^{\frac{2}{\alpha }}}{\left\| {{\bf{s}} - {{\bf{X}}_i}} \right\|^2}, ~\forall i \in {\cal{I}}, j \in {\cal{J}}_i,\label{5-3}\\
			& \tau_i \ge \frac{1}{L_i}, ~\forall i \in \cal{I}, \label{5-4}\\
			& \text{(\ref{4-3}), (\ref{4-5}),}
		\end{align}
	\end{subequations}
	where (\ref{5-3}) is derived from (\ref{4-2}), since $\|\mathbf{s}-\mathbf{X}_j\|^2$ is convex with respect to $\mathbf{s}$, while the left-hand-side of (\ref{5-3}) is its lower bound obtained by using the first-order Taylor expansion at a local point $\mathbf{\tilde s}$. However, the problem is still non-convex due to the non-convexity of constraints (\ref{5-4}) and (\ref{4-3}).
	To address the non-convex constraint (\ref{5-4}), we introduce the variables $ \phi  $, $ \varphi_i $ as
	\begin{equation}
		\phi  \ge {d_{B - S}}, \label{059a}
	\end{equation}
	\begin{equation}
		\varphi_i  \ge {d_{S - {U_i}}}, ~\forall i \in \cal {I},\label{059b}
	\end{equation}
	and the constraint (\ref{5-4}) can be replaced by
	\begin{equation} 
		\tau_i^{\frac{1}{\alpha}} \ge \phi\varphi_i,~\forall i \in \cal {I}, \label{059c}
	\end{equation}
	where the right-hand-side can be reformulated as $\phi\varphi_i \buildrel \Delta \over = \frac{{{{\left( {\phi  + \varphi_i } \right)}^2}}}{2}-\frac{\phi^2-\varphi_i ^2}{2}$. Then by obtaining its lower bound at the local point $(\tilde \phi, \tilde \varphi_i )$, (\ref{5-4}) can be finally replaced by
	
	\begin{equation}
		\tau_i^{\frac{1}{\alpha}} \ge \frac{{{{\left( {\phi  + \varphi_i } \right)}^2}}}{2} - \frac{{{{\tilde \phi }^2} - {{\tilde \varphi_i }^2}}}{2} - \tilde \phi \left( {\phi  - \tilde \phi } \right) - \tilde \varphi_i \left( {\varphi_i  - \tilde \varphi_i } \right),~\forall i \in \cal {I}.
		\label{060}
	\end{equation}
	Furthermore, the constraint (\ref{4-3}) can be reformulated as
	\begin{equation}
		{|\mathbf{r}_k^\dag\boldsymbol{\Phi}_{k,p}\mathbf{G}\mathbf{w}_i|^2\left(\sum\limits_{j \in {\cal{J}}_i}u_{i,j}+\tau_i\sigma_i^2\right)\ge u_{i,i}\left({\sum\limits_{j \in {\cal{J}}_i}|\mathbf{r}_k^\dag\boldsymbol{\Phi}_{k,p}\mathbf{G}\mathbf{w}_j|^2}+\tau_k\sigma_k^2\right)}, \forall i\in {\cal{I}}, j,k \in {\cal{J}}_i.
		\label{29}
	\end{equation}
	Finally, the optimization problem can be rewritten as
	\begin{subequations} \label{pp5}
		\begin{align}
			\mathop {\max }\limits_{\boldsymbol{\Phi}_s }~&\sum\limits_{i = 1}^K {{\eta _i}}  R_i^s
			\label{p5-1}\\
			{\text{s.t.}}\quad
			& {\text{(\ref{4-5}), (\ref{5-2}), (\ref{059a})-(\ref{29}),}}
		\end{align}
	\end{subequations}
	where $\boldsymbol{\Phi}_s \triangleq \{\mathbf{s}, \phi, \{\varphi_i\}, \{\tau_i\}\}$. The STAR-RIS location optimization algorithm is detailed in {\bf{Algorithm 2}}.
	\begin{algorithm}
		\LinesNumbered
		\caption{Proposed Algorithm for Optimizing the Location in the $\kappa$-th AO Iteration.}
		\KwIn { The obtained active BF $ {\bf{w}}_i^{(\kappa-1)} $, and the passive BF $ {\bf{v}}_t^{(\kappa-1)} $ and $  {\bf{v}}_r^{(\kappa-1)} $, and the initial feasible points $ \boldsymbol{\tilde \Phi}_s^{(0)} $.}
		Solve (\ref{pp5}) in the predefined deployment region to obtain an optimized location $ \mathbf{s}^{{ini}}$\;
		{\bf{Repeat}} \;
		Solve (\ref{pp5}) in a small region near the $\mathbf{s}^{ini}$ to obtain the optimized location $ \mathbf{s}^{(\kappa_s)} $\;
		Update $ \boldsymbol{\tilde \Phi}_s^{(\kappa_s)}$, calculate the objective value $ F^{(\kappa_s)}$, $\kappa_s = \kappa_s+1 $\;
		{\bf{Until:}} $ \left|F^{(\kappa)}-F^{(\kappa-1)}\right| \le  \delta $\;
		{\KwOut{ ${\bf{s}}^{(\kappa)} $.}}
	\end{algorithm}
	
	Based on the above sub-problems and their optimization algorithms, an AO based algorithm is proposed to alternately obtain the active BF at the BS, the passive BF as well as the location of the STAR-RIS. The details of our proposed AO based algorithm are summarized in {\bf{Algorithm 3}}.
	\begin{algorithm}
		\label{algo3}
		\LinesNumbered
		\caption{Proposed AO based Algorithm for the Beamformer-based STAR-RIS-NOMA.}
		\KwIn {Initiailizing Locations of users and the BS, the channel vectors $ \mathbf{r}_i $ and $\mathbf{g}$.}
		Initialize the passive BF $ \mathbf{v}_p^{(0)} $, the active BF $ \mathbf{w}_i^{(0)} $, and location of the STAR-RIS $ \mathbf{s}^{(0)} $\;
		Set the tolerance of iteration accuracy $ \tilde \delta $\;
		{\bf{Repeat}} \;
		Update $ {\bf{w}}_i^{(\kappa)}$ by solving (\ref{p6}) with $ {\bf{v}}_t^{(\kappa-1)} $, $ {\bf{v}}_r^{(\kappa-1)} $, and $\mathbf{s}^{(\kappa-1)}$\;
		Update $ {\bf{v}}_t^{(\kappa)} $ and $ {\bf{v}}_r^{(\kappa)} $ by solving (\ref{p16}) with obtained $\mathbf{w}_i^{(\kappa-1)}$ and $\mathbf{s}^{(\kappa-1)}$\;
		Update $ {\bf{s}}^{(\kappa)} $ by {\bf{Algorithm 2}} with $ {\bf{v}}_t^{(\kappa-1)} $, $ {\bf{v}}_r^{(\kappa-1)} $ and $\mathbf{w}_i^{(\kappa-1)}$\;
		Calculate the objective value $ F^{(\kappa)} $\;
		{\bf{Until:}} $ \left|F^{(\kappa)}-F^{(\kappa-1)}\right| \le \tilde \delta $\;
		{\KwOut{Optimized $ \mathbf{v}_t $, $ \mathbf{v}_r $, $ {\bf{w}}_i $, and ${\bf{s}} $.}}
	\end{algorithm}
	\vspace{-30pt}
	\subsection{Convergence Analysis}
	\label{convergence1}
	In our AO based algorithm, the solutions obtained by solving (\ref{p6}), (\ref{p16}), and (\ref{pp5}) are used as the input of the other sub-problems, and the convergence can be proved\cite{mu2021joint}.
	We first define $ F\left(\{\mathbf{w}_i^{(\kappa)}\}, \mathbf{v}_p^{(\kappa)}, \mathbf{s}^{(\kappa)}\right) $ as the objective value of (\ref{p5}) after the $ \kappa $-th iteration, which may be formulated as:
	\begin{equation} \label{eq34}
		\begin{aligned}
			F\left( {\{ {\bf{w}}_i^{(\kappa )}\} ,{\bf{v}}_p^{(\kappa )},{{\bf{s}}^{(\kappa )}}} \right) & \mathop  = \limits^{(a)} F_{\{ {\bf{w}}_i^{(\kappa )}\} }^{lb}\left( {\{ {\bf{w}}_i^{(\kappa )}\} ,{\bf{v}}_p^{(\kappa )},{{\bf{s}}^{(\kappa )}}} \right) \mathop  \le \limits^{(b)} F_{\{ {\bf{w}}_i^{(\kappa )}\} }^{lb}\left( {\{ {\bf{w}}_i^{(\kappa  + 1)}\} ,{\bf{v}}_p^{(\kappa )},{{\bf{s}}^{(\kappa )}}} \right)\\& \mathop  \le \limits^{(c)} F\left( {\{ {\bf{w}}_i^{(\kappa  + 1)}\} ,{\bf{v}}_p^{(\kappa )},{{\bf{s}}^{(\kappa )}}} \right),
		\end{aligned}
	\end{equation}
	where $ F_{\{ {\bf{w}}_i^{(\kappa )}\} }^{lb} \left(\cdot\right) $ denotes the objective value of (\ref{p6}) with the obtained $ \{\mathbf{w}_i^{(\kappa)}\} $. In \eqref{eq34}, we have $ (a) $ follows from the fact that the first-order Taylor expansion in (\ref{p11}) is tight, $( b) $ holds when $ \{\mathbf{w}_i^{(\kappa+1)}\} $ is optimized, and $ (c) $ is satisfied because (\ref{p6}) is always the lower bound. Hence, the original problem is non-decreasing, which can be expressed as
	\begin{equation}
		\begin{aligned}
			F\left( {\{ {\bf{w}}_i^{(\kappa )}\} ,{\bf{v}}_p^{(\kappa )},{{\bf{s}}^{(\kappa )}}} \right) & =  F_{\{ {\bf{v}}_p^{(\kappa )}\} }^{lb}\left( {\{ {\bf{w}}_i^{(\kappa )}\} ,{\bf{v}}_p^{(\kappa )},{{\bf{s}}^{(\kappa )}}} \right) \le F_{\{ {\bf{v}}_p^{(\kappa )}\} }^{lb}\left( {\{ {\bf{w}}_i^{(\kappa )}\} ,{\bf{v}}_p^{(\kappa  + 1)},{{\bf{s}}^{(\kappa )}}} \right) \\& \le F\left( {\{ {\bf{w}}_i^{(\kappa )}\} ,{\bf{v}}_p^{(\kappa  + 1)},{{\bf{s}}^{(\kappa )}}} \right),
		\end{aligned}
	\end{equation}
	and
	\begin{equation}
		\begin{aligned}
			F\left( {\{ {\bf{w}}_i^{(\kappa )}\} ,{\bf{v}}_p^{(\kappa )},{{\bf{s}}^{(\kappa )}}} \right) &= F_{\{ {{\bf{s}}^{(\kappa )}}\} }^{lb}\left( {\{ {\bf{w}}_i^{(\kappa )}\} ,{\bf{v}}_p^{(\kappa )},{{\bf{s}}^{(\kappa )}}} \right) \le F_{\{ {{\bf{s}}^{(\kappa )}}\} }^{lb}\left( {\{ {\bf{w}}_i^{(\kappa )}\} ,{\bf{v}}_p^{(\kappa )},{{\bf{s}}^{(\kappa  + 1)}}} \right) \\& \le F\left( {\{ {\bf{w}}_i^{(\kappa )}\} ,{\bf{v}}_p^{(\kappa )},{{\bf{s}}^{(\kappa  + 1)}}} \right).
		\end{aligned}
	\end{equation}
	Finally, we arrive at
	\begin{equation}
		F\left( {\{ {\bf{w}}_i^{(\kappa )}\} ,{\bf{v}}_p^{(\kappa )},{{\bf{s}}^{(\kappa )}}} \right) \le F\left( {\{ {\bf{w}}_i^{(\kappa  + 1)}\} ,{\bf{v}}_p^{(\kappa  + 1)},{{\bf{s}}^{(\kappa  + 1)}}} \right).
		\label{p29}
	\end{equation}
	
	\begin{remark}
		Equation (\ref{p29}) indicates that the objective value of the original problem is non-decreasing in each iteration of {\bf{Algorithm 3}}, and our proposed algorithm is guaranteed to converge due to the finite value of the WSR.
	\end{remark}
	\vspace{-20pt}
	\subsection{Complexity Analysis}
	\label{complexity1}
	Observe that the total complexity of the AO based algorithm depends on the complexity of each sub-problem. The complexity of solving (\ref{p6}) is $	O_1 = {\cal{O}}\left(I^{\max}_1\max\left(N_t, (2N+1)^4\sqrt{N_t}\log_2\frac{1}{\tilde \delta_1}\right)\right)$, the complexity of solving (\ref{p16}) is $O_2= {\cal{O}}\left(I^{\max}_2\max\left(M, (2N)^4\sqrt{M}\log_2\frac{1}{\tilde \delta_2}\right)\right)$, where $\delta_1$ and $\delta_2$ represent the solution accuracy, while $I_1^{\max}$ and $I_1^{\max}$ denote the number of iterations. The complexity of the position optimization is $O_3 = {\cal{O}}\left(2N+4\right)^{3.5}$. The total complexity of our proposed AO based algorithm is ${\cal{O}}\left[I_{\max}^{AO}\left(O_1+O_2+O_3\right)\right]$, where $ I_{\max}^{AO} $ is the number of iterations in {\bf{Algorithm 3}}. Specifically, the complexity of exhaustively searching all possible combinations is ${\cal{O}}\left(N!\right)$.
	\vspace{-25pt}
	\section{Proposed Algorithm for Cluster-based NOMA }
	\label{section4}
	In this section, an AO based algorithm is invoked by decomposing the original problem (\ref{p40}) into several sub-problems to address the coupled optimization variables, where the PAFs and active BF at the BS as well as the passive BF and the location of the STAR-RIS are alternately optimized. Furthermore, a low-complexity decoding order design based on the equivalent-channel-gain of the users is employed  \cite{zuo2021joint}.
	\begin{remark}
		Consider the decoding order of any users $k$ and $i$ in any cluster. Then their equivalent-channel-gain is used to decide the decoding order, where if 
		\begin{equation}
			{	\frac{L_{n,k}|\mathbf{r}_{n,k}^\dag \boldsymbol{\Phi}_{n,p}\mathbf{G}\mathbf{w}_n|^2}{L_{n,k}\sum\limits_{n'\ne n}|\mathbf{r}_{n,k}^\dag \boldsymbol{\Phi}_{n,p}\mathbf{G}\mathbf{w}_{n'}|^2+\sigma_{n,k}^2} \ge \frac{L_{n,i}|\mathbf{r}_{n,i}^\dag \boldsymbol{\Phi}_{n,p}\mathbf{G}\mathbf{w}_n|^2}{L_{n,i}\sum\limits_{n'\ne n}|\mathbf{r}_{n,i}^\dag \boldsymbol{\Phi}_{n,p}\mathbf{G}\mathbf{w}_{n'}|^2+\sigma_{n,i}^2}},
			\label{p41}
		\end{equation}
		we have $\pi(n,k) > \pi(n,i)$.
	\end{remark}
	
	\begin{remark}
		It may be readily shown that when we use the equivalent channel gain for determining the decoding order, the constraint (\ref{7-4}) can be satisfied \cite{zuo2021joint}. However, the equivalent-channel-gain of users is related to both the active and passive BF as well as to the location. Hence the decoding order has to be updated after each iteration.
	\end{remark}
	
	According to {\bf{Remark 4}}, (\ref{7-4}) can be released by employing the detect order in (\ref{p41}), while the optimization problem (\ref{p40}) is still non-convex due to the non-convex objective function as well as constraints. Hence, we decompose the original problem (\ref{p40}) into sub-problems and solve them iteratively as follows.
	\vspace{-20pt}
	\subsection{The Sub-problem of Active BF and PAFs Design}
	\label{4A}
	By denoting $ {{\bf{\bar h}}_{n,i}} = \sqrt{L_{n,i}} {{\bf{r}}_{n,i}^\dag{\mathbf{\Theta} _{n,p}}{\bf{G}}} $, $ {{{\bf{\bar H}}}_{n,i}} = {\bf{\bar h}}_{n,i}^\dag{\bf{\bar h}}_{n,i} $, and $\mathbf{W}_n = \mathbf{w}_n\mathbf{w}_n^\dag$ to employ the SDP method, the optimization problem can be formulated as
	\begin{subequations}\label{p35}
		\begin{align}
			\mathop {\max }\limits_{{p_{n,i}},{\bf{W}}_n} &\sum\limits_{n=1}^{N_c}\sum\limits_{i = 1}^{K} {{\eta _{n,i}}} {\log _2}\left( {\frac{{(p_{n,i}+\sum\limits_{j \in {{\cal{J}}_{n,i}}}{p_{n,j}}) {\rm{Tr}}\left( {{\bf{W}}_n{{{\bf{\bar H}}}_{n,i}}} \right) + \sum\limits_{n' \ne n}{\rm{Tr}}\left(\mathbf{W}_{n'}\mathbf{\bar H}_{n,i}\right)+ \sigma _{n,i}^2}}{{\sum\limits_{j \in {\cal{J}}_{n,i}} {{p_{n,j}}{\rm{Tr}}\left( {{\bf{W}}_n{{{\bf{\bar H}}}_{n,i}}} \right)+\sum\limits_{n' \ne n}{\rm{Tr}}\left(\mathbf{W}_{n'}\mathbf{\bar H}_{n,i}\right) + \sigma _{n,i}^2} }}} \right)
			\label{8-1}\\
			{\text{s.t.}}\quad
			& \mathbf{W}_n \succeq 0, 
			\label{8-2}\\
			&{(p_{n,i}-\gamma_{n,i}\sum\limits_{j \in {\cal J}_{n,i}}p_{n,j})}{\rm{Tr}}\left(\mathbf{W}_n\mathbf{\bar H}_{n,i}\right)-\gamma_{n,i}\sum\limits_{n' \ne n}{\rm{Tr}}\left(\mathbf{W}_n\mathbf{\bar H}_{n,i}\right)-\gamma_{n,i}\sigma^2_{n,i} \ge 0,
			\label{p8-3}\\
			& \sum\limits_{i = 1}^{K} {{p_{n,i}}}  = 1, 
			\label{8-6} \\
			& \sum\limits_{n=1}^{N_c}{\rm{Tr}} \left( {\bf{W}}_n \right) \le P_{\max},
			\label{8-7}\\
			& {\rm{Rank}} \left( \mathbf{W}_n \right) =1, 
			\label{8-3}
		\end{align}
	\end{subequations}
	where (\ref{p8-3}) is equivalent to (\ref{7-3}), while $\gamma_{n,i} = 2^{R_{n,{i \to i}}^{\min}}-1$ denotes the minimal SINR of the $i $-th user in the $n$-th cluster.
	To tackle the non-convexity of the objective function (\ref{8-1}), we introduce the variables $ \{ x_{n,i}\} $ and $ \{ y_{n,i}\} $ as
	\begin{equation}
		{e^{{{ x}_{n,i}}}} \le (p_{n,i}+\sum\limits_{j \in {\cal{J}}_{n,i}} {{p_{n,j}}}) {\rm{Tr}}\left( {{\bf{W}}_n{{{\bf{\bar H}}}_{n,i}}} \right) + \sum\limits_{n' \ne n}{\rm{Tr}}\left(\mathbf{W}_{n'}\mathbf{\bar H}_{n,i}\right) + \sigma _{n,i}^2,
		\label{38}
	\end{equation}
	\begin{equation}
		{e^{{y_{n,i}}}} \ge \sum\limits_{j \in {\cal{J}}_{n,i}} {{p_{n,j}}{\rm{Tr}}\left( {{\bf{W}}_n{{{\bf{\bar H}}}_{n,i}}} \right) + \sum\limits_{n' \ne n}{\rm{Tr}}\left(\mathbf{W}_{n'}\mathbf{\bar H}_{n,i}\right) + \sigma _{n,i}^2}.
		\label{39}
	\end{equation}
	Then the optimization problem can be formulated as
	\begin{subequations}
		\begin{align}
			\mathop {\max }\limits_{\scriptstyle{p_{n,i,}}{\mathbf{W}_n}\hfill\atop
				\scriptstyle{x_{n,i}},{y_{n,i}}\hfill}&{\log _2}e\sum\limits_{n=1}^N\sum\limits_{i = 1}^{K_n} {{\eta _{n,i}}} \left( {{x_{n,i}} - {y_{n,i}}} \right)
			\label{p9-1}\\
			{\text{s.t.}}\quad
			& {\text{(\ref{8-2})-(\ref{8-3}), (\ref{38}), (\ref{39})}}.
		\end{align}
	\end{subequations}
	
	Now, we turn our attention to the transform action of the non-convex constraint (\ref{38}) by introducing the variables $ \{\varsigma_{n,i}\} $ as
	\begin{equation}
		\left[ {\begin{array}{*{20}{c}}
				{\sum\limits_{j \in {\cal{J}}_i} {{p_{n,j}}} + p_{n,i} }&{{\varsigma _{n,i}}}\\
				{{\varsigma _{n,i}}}&{{\rm{Tr}}\left( {{\bf{W}}_n{{{\bf{\bar H}}}_{n,i}}} \right)}
		\end{array}} \right] \succeq 0.
		\label{46}
	\end{equation}
	Then (\ref{38}) can be reformulated by using the first-order Taylor expansion of $\varsigma^2_{n,i}$ at the feasible point $\tilde \varsigma_{n,i}$ as
	\begin{equation}
		2\tilde \varsigma _{n,i}{\varsigma _{n,i}} - {{\tilde \varsigma _{n,i}}^2} + \sum\limits_{n' \ne n}{\rm{Tr}}\left(\mathbf{W}_{n'}\mathbf{\bar H}_{n,i}\right)+ \sigma _{n,i}^2 \ge e^{x_{n,i}}.
		\label{45}
	\end{equation}
	As for the non-convex constraint (\ref{39}), we introduce the variables $\{c_{n,i}\}$ and $\{d_{n,i}\}$ satisfying
	\begin{equation}
		{c_{n,i}} \ge {\rm{Tr}}\left( {{\bf{W}}_n{{{\bf{\bar H}}}_{n,i}}} \right),
		\label{43}
	\end{equation}
	\begin{equation}\label{41b}
		1 \ge {d_{n,i}} \ge \sum\limits_{j \in {\cal{J}}_{n,i}} p_{n,j},
	\end{equation}
	where we have
	\begin{equation}\label{41c}
		\sum\limits_{j \in {J_{n,i}}} {{p_{n,j}}} {\rm{Tr}}\left( {{{\bf{W}}_n}{{{\bf{\bar H}}}_{n,i}}} \right) 
		\mathop  \le 
		\frac{{{{({c_{n,i}} + {d_{n,i}})}^2}}}{2} - \frac{{\tilde c_{n,i}^2 - \tilde d_{n,i}^2}}{2} - {{\tilde c}_{n,i}}({c_{n,i}} - {{\tilde c}_{n,i}}) - {{\tilde d}_{n,i}}({d_{n,i}} - {{\tilde d}_{n,i}}).
	\end{equation}
	The right-hand-side of (\ref{41c}) follows from the first order Taylor expansion at a local point $ (\tilde c_{n,i}, \tilde d_{n,i}) $. Finally, constraint (\ref{39}) can be replaced by
	\begin{equation}\label{41d}
		\begin{aligned}
			&\frac{{{{({c_{n,i}} + {d_{n,i}})}^2}}}{2} - \frac{{\tilde c_{n,i}^2 - \tilde d_{n,i}^2}}{2} - {{\tilde c}_{n,i}}({c_{n,i}} - {{\tilde c}_{n,i}}) - {{\tilde d}_{n,i}}({d_{n,i}} - {{\tilde d}_{n,i}}) +  \sum\limits_{n' \ne n}{\rm{Tr}}\left(\mathbf{W}_{n'}\mathbf{\bar H}_{n,i}\right) + \sigma _{n,i}^2 \\& \le e^{\tilde y_{n,i}} (y_{n,i}-\tilde y_{n,i}+1),
		\end{aligned}
	\end{equation}
	where $e^{\tilde y_{n,i}} (y_{n,i}-\tilde y_{n,i}+1)$ is the lower bound of $ e^{y_{n,i}} $ at the local point $\tilde y_{n,i}$ due to the convexity of $ e^{y_{n,i}} $.
	
	Furthermore, the optimization problem after relaxing the rank-one constraint can be finally formulated as
	\begin{subequations} \label{p46}
		\begin{align}
			\mathop {\max }\limits_{\mathbf{\Phi}}~ 	&{\log _2}e\sum\limits_{n=1}^N\sum\limits_{i = 1}^{K_n} {{\eta _{n,i}}} \left( {{x_{n,i}} - {y_{n,i}}} \right)
			\label{p10-1}\\
			{\text{s.t.}}\quad
			& {\text{(\ref{8-2})-(\ref{8-7}), (\ref{43})-(\ref{45}),}}
		\end{align}
	\end{subequations}
	where ${\boldsymbol{\Phi}} = \{\{ {p_{n,i}}\} ,{\bf{W}}_n,\{ {x_{n,i}}\} ,\{ {y_{n,i}}\}, {\{c_{n,i}\}},{\{d_{n,i}\}},{\{\varsigma _{n,i}\}}\}$.Now, problem (\ref{p46}) is convex and can be efficiently solved by CVX. The detailed procedure is given as {\bf{Algorithm 4}}.
	\begin{algorithm}
		\LinesNumbered
		\caption{Algorithm for Optimizing the Active BF and the PAFs in the $\tilde \kappa$-th AO Iteration.}
		\KwIn {Obtained $ \mathbf{v}_t^{(\tilde \kappa-1)} $, $ \mathbf{v}_r^{(\tilde \kappa-1)} $ and $\mathbf{s}^{(\tilde \kappa-1)}$.}
		Set the tolerance of iteration accuracy $ \tilde \delta $, max iteration time $ I_{max} $, initial iteration number $ \tilde \kappa_b=0 $, and feasible point set $ {\bf{\tilde \Phi}}^{(0)} $\;
		\While{$\tilde \kappa_b \le I_{max}$ and $ \delta > \tilde \delta $}{
			$ \tilde \kappa_b = \tilde \kappa_b+1 $\;
			Solve (\ref{p46}) to obtain the solution $ \mathbf{W}_n^{(\tilde \kappa_b)}$ and $ \{p_{n,i}^{(\tilde \kappa_b)} \}$ \;
			Update  $ {\bf{\tilde \Phi}}^{(\tilde \kappa_b)} $, where $ \tilde y_{n,i}^{\left( \kappa_b \right)} = \ln \left( \sum\limits_{j = i + 1}^{K_n} {{p_{n,j}^{(\kappa_b)}}{\rm{Tr}}\left( {{\bf{W}}_n^{(\kappa_b)}{{{\bf{\bar H}}}_{n,i}}} \right) + \sum\limits_{n' \ne n}{\rm{Tr}}\left(\mathbf{W}_{n'}^{(\kappa_b)}\mathbf{\bar H}_{n,i}\right) + \sigma _{n,i}^2}\right) $
			\;
			Update objective value $ F^{(\tilde \kappa_b)} $ and calculate $ \delta = | F^{(\kappa_b)}-F^{(\kappa_b-1)}|$\;
		}
		\KwOut {Optimal $ {\bf{W}}_n^{(\tilde \kappa)} $ and $ p_{n,i}^{(\tilde \kappa)} $.}
	\end{algorithm}
	\vspace{-35pt}
	\subsection{The Sub-problem of Passive BF Design }
	Similar to the procedure of Section \ref{4A}, to tackle the non-convexity of the problem, we define $ \mathbf{V}_{n,p} = \mathbf{v}_{n,p}^\dag\mathbf{v}_{n,p} $, which satisfies $\mathbf{V}_p \succeq 0$ and $\text{Rank}\left(\mathbf{V}_p\right)=1,~p \in \{t,r\}$. Then the problem of optimizing $ \mathbf{v}_p $ for a given $\mathbf{w}_n$ and $\mathbf{s}$ can be formulated as
	\begin{subequations} \label{p56}
		\begin{align}
			\mathop  {\max }\limits_{\scriptstyle{{\bf{V}}_t},{{\bf{V}}_r},\hfill\atop
				A_{n,i}^c,B_{n,i}^c\hfill}{\rm{ }}~ &\sum\limits_{n = 1}^{N_c} {\sum\limits_{i = 1}^{{K}} {{\eta _{n,i}}} }{\log _2}\left( {1 + \frac{1}{{{{\tilde A}_{n,i}^{c}}{{\tilde B}_{n,i}^{c}}}}} \right) - \frac{{{{\log }_2}e\left( {{A^c_{n,i}} - {{\tilde A}_{n,i}^{c}}} \right)}}{{{({\tilde A}_{n,i}^{c})}^2{{\tilde B}_{n,i}^{c}} + {{\tilde A}_{n,i}^{c}}}} - \frac{{{{\log }_2}e\left( {{B^c_{n,i}} - {{\tilde B}_{n,i}^{c}}} \right)}}{{{{\tilde A}_{n,i}^{c}}{({\tilde B}_{n,i}^{c})}^2 + {{\tilde B}_{n,i}^{c}}}}
			\label{p11-1}\\
			{\text{s.t.}}\quad
			& \frac{1}{A^c_{n,i}} \le  {p_{n,i}}{\rm{Tr}}\left( {{{\bf{V}}_{n,p}}{{{\bf{\hat H}}}_{n,i}}} \right),
			\label{11-2}\\
			& B^c_{n,i} \ge {{\sum\limits_{j \in {{\cal J}_{n,i}}} {{p_{n,j}}{\rm{Tr}}\left( {{{\bf{V}}_{n,p}}{{{\bf{\hat H}}}_{n,i}}} \right) + \sum\limits_{n' \ne n} {{\rm{Tr}}} \left( {{{\bf{V}}_{n,p}}{{{\bf{\hat H}}}_{n,n',i}}} \right) + \sigma _{n,i}^2} }},
			\label{11-10}\\
			&	{{(p_{n,i}-\gamma_{n,i}\sum\limits_{j \in {\cal J}_{n,i}}p_{n,j})}{\rm{Tr}}\left(\mathbf{V}_{n,p}\mathbf{\hat H}_{n,i}\right)-\gamma_{n,i}(\sum\limits_{n' \ne n}{\rm{Tr}}\left(\mathbf{V}_{n,p}\mathbf{\hat H}_{n,n',i}\right)+\sigma^2_{n,i}) \ge 0,}
			\label{11-7}\\
			& {\text{(\ref{p5-5})-(\ref{p5-8})}},
		\end{align}
	\end{subequations}
	where $ \mathbf{\hat h}_{n,i} = \sqrt{L_{n,i}}{\rm{diag}}( \mathbf{r}_{n,i}^\dag)\mathbf{G}\mathbf{w}_n$, $ \mathbf{\hat h}_{n,n',i} = \sqrt{L_{n,i}}{\rm{diag}}( \mathbf{r}_{n,i}^\dag)\mathbf{G}\mathbf{w}_{n'}$, $ \mathbf{\hat H}_{n,i} = \mathbf{\hat h}_{n,i}\mathbf{\hat h}_{n,i}^\dag $, and $ \mathbf{\hat H}_{n,n',i} = \mathbf{\hat h}_{n,n',i}\mathbf{\hat h}_{n,n',i}^\dag $, $A_{n,i}^c$ and $B_{n,i}^c$ are the slack variables introduced and (\ref{p11-1}) is the lower bound of $ \sum_{n=1}^{N_c}\sum_{i=1}^{K}\tilde R_{n,i \to i} $.
	It may be readily seen that problem (\ref{p56}) is similar to problem (\ref{p16}), hence we omit the detailed algorithm for brevity.
	\vspace{-30pt}
	\subsection{The Sub-problem of Deployment Location Design}
	Similar to (\ref{pp5}), the deployment location sub-problem can be formulated as
	\begin{subequations} \label{pp51}
		\begin{align}
			\mathop {\max }\limits_{\bf{s}} ~&{\sum_{n=1}^{N_c}}{\sum_{i=1}^{K}} {\eta_{n,i}}\tilde R_{n,i \to i}
			\label{51-1}\\
			{\text{s.t.}}\quad
			& \mathbf{s}\left(x,y\right) \in \cal{X}\times\cal{Y},
			\label{51-2}\\
			& \tilde R_{n,{i\to i}} \ge R^{\min}_{n,{i \to i}}
			\label{51-3}.
		\end{align}
	\end{subequations}
	To address the non-convexity objective function in (\ref{51-1}), we introduce
	\begin{equation} \label{r52}
		\tau_{n,i} \ge \frac{1}{L_{n,i}},
	\end{equation}
	and the linearized lower bound of $ \tilde R_{n,i \to i} $ is given by employing the first-order Taylor expansion at the local point $\tilde \tau_{n,i}$ as
	
	\begin{equation}
		\begin{aligned}
			Z_{n,i} =   &{{\log }_2}\left( {1 + \frac{{{\tilde u_{n,i}}}}{{\sum\limits_{j \in {\cal{J}}_{n,i}} {{\tilde u_{n,j}}}  +u_{n,i,n'}+ \tilde \tau _{n,i}\sigma _i^2}}} \right)\\& - \frac{{{\tilde u_{n,i}}\sigma _i^2{{\log }_2}e\left( {\tau _{n,i}}-{\tilde \tau _{n,i}} \right)}}{{\left( \tilde u_{n,i}+{\sum\limits_{j \in {\cal{J}}_{n,i}}\tilde u_{n,j}+ \tilde u_{n,i,n'} +\tilde \tau _{n,i}\sigma _{n,i}^2} \right)\left( {\sum\limits_{j \in {\cal{J}}_{n,i}}\tilde u_{n,j}+ \tilde u_{n,i,n'} +\tilde \tau _{n,i}\sigma _{n,i}^2}  \right)}},
		\end{aligned}
	\end{equation}
	where ${{\tilde u}_{n,i}} = {{p_{n,i}}} {| {{\bf{r}}_{n,i}^\dag{{\bf{\Theta }}_{n,p}}{\bf{G}}\mathbf{w}_n} |^2}$, ${{\tilde u}_{n,i,n'}} = {| {{\bf{r}}_{n,i}^\dag{{\bf{\Theta }}_{n,p}}{\bf{G}}\mathbf{w}_{n'}} |^2}$.
	Similar to (\ref{059a}), (\ref{059b}) and (\ref{060}), by introducing 
	\begin{equation}
		\varphi_{n,i}  \ge {d_{S - {U_{n,i}}}},\label{p48}
	\end{equation}
	the non-convex constraint (\ref{r52}) can be replaced by
	\begin{equation}
		{ \tau_{n,i} ^{\frac{1}{\alpha }}}  \ge   \frac{{{{\left( {\phi  + \varphi_{n,i} } \right)}^2}}}{2} - \frac{{{{\tilde \phi}^2} + {{\tilde \varphi_{n,i}}^2}}}{2} - \tilde \phi \left( {\phi  - \tilde \phi} \right)  - \tilde \varphi_{n,i} \left( {\varphi_{n,i} - \tilde \varphi_{n,i}} \right),
		\label{p49}
	\end{equation}
	where the definition of $\phi$ is the same as in (\ref{059a}). 
	
	Finally, problem (\ref{pp51}) can be equivalently written as
	\begin{subequations} \label{p61}
		\begin{align}
			\mathop {\max }\limits_{\mathbf{s}, \tau_i }~&\sum\limits_{n = 1}^{N_c}\sum\limits_{i = 1}^{K} {{\eta _{n,i}}}Z_{n,i}
			\label{13-1}\\
			{\text{s.t.}}\quad
			&\mathbf{s}\left(x,y\right) \in \cal{X}\times \cal{Y},
			\label{13-2}\\
			&{{{{\tilde u_{n,i}}}}-\gamma_{n,i}\left({{\sum\limits_{j \in {\cal{J}}_{n,i}} {{\tilde u_{n,j}}}  +u_{n,i,n'}+ \tau _{n,i}\sigma _i^2}}\right) \ge 0,}
			\label{13-3}\\
			& Z_{n,{i}} \ge R^{\min}_{n,{i \to i}},\\
			& {\text{(\ref{059a}), (\ref{p48}), (\ref{p49}),}}
		\end{align}
	\end{subequations}
	where (\ref{13-3}) is equivalent to (\ref{7-3}).
	
	Since the optimization problem (\ref{p61}) is similar to problem (\ref{pp5}), the procedure of solving it is similar to {\bf{Algorithm 2}}. By Recalling the procedure given in {\bf{Algorithm 3}}, the AO based algorithm of obtaining the PAFs and the active BF at the BS as well as the passive BF and the location of the STAR-RIS can be formulated by alternately solving problems (\ref{p46}), (\ref{p56}), and (\ref{p61}). However, the detailed algorithm is omitted here for brevity. 
	
	\vspace{-15pt}
	\subsection{Convergence Analysis}
	
	Let us now consider the sub-problem (\ref{p35}) and define the solution obtained at the $  \tilde \kappa_b $-th iteration as $\mathbf {\Phi}^{(\tilde \kappa_b)} = \{{\bf{ W}}_n^{(\tilde \kappa_b)}, p_{n,i}^{(\tilde \kappa_b)},  x_{n,i}^{(\tilde \kappa_b)}, y_{n,i}^{(\tilde \kappa_b)},  a_{n,i}^{(\tilde \kappa_b)},  b_{n,i}^{(\tilde \kappa_b)},  \varsigma _{n,i}^{(\tilde \kappa_b)}\} $. Note that $  y_{n,i}^{(\tilde \kappa_b)} $ is the feasible solution of (\ref{p35}). Hence it may be readily shown that the solution $ {\tilde y_{n,i}}^{(\tilde \kappa_b)} $ obtained must satisfy $ {\tilde y_{n,i}}^{(\tilde \kappa_b)} \le y_{n,i}^{(\tilde \kappa_b)}$  due to the specific form of the objective function in (\ref{p35}). Then, according to \cite{lin2019joint}, we have
	\begin{equation}
		{e^{y_{n,i}^{\left( {\tilde \kappa_b  + 1} \right)}}} = \sum\limits_{j = i + 1}^{K_n} {p_{n,j}^{(\tilde \kappa_b )}{\rm{Tr}}\left( {{{\bf{W}}_n^{(\tilde \kappa_b )}}{{{\bf{\bar H}}}_{n,i}}} \right) + \sigma_{n,i}^2}  = {e^{\tilde y_{n,i}^{\left( \tilde \kappa_b  \right)}}}\left( {\tilde y_{n,i}^{\left( \tilde \kappa_b  \right)} - y_{n,i}^{\left( \tilde \kappa_b  \right)} + 1} \right) \mathop  \le \limits^{(d )} {e^{\tilde y_{n,i}^{\left( \tilde \kappa_b  \right)}}},
		\label{p50}
	\end{equation}
	where $ (d )$ holds due to the first-order Taylor expansion. Hence it is readily seen that
	$ 	y_{n,i}^{(\tilde \kappa_b+1)} \le \tilde y_{n,i}^{(\tilde \kappa_b)} \le y_{n,i}^{(\tilde \kappa_b)} $.
	Furthermore, we have
	$ \sigma _{n,i}^2 \le {y_{n,i}^{(\tilde \kappa_b)}} \le \infty $. Therefore, $ y_{n,i}^{(\tilde \kappa_b)} $ is a bounded monotonic function, which converges as the iteration index $ \tilde \kappa_b $ increases. Meanwhile, $ \tilde y_{n,i}^{(\tilde \kappa_b)} $ also converges due to the convergence of $ y_{n,i}^{(\tilde \kappa_b)} $.
	Therefore, the objective value obtained converges to a stable value.
	
	\begin{remark}
		Similarly, the convergence of our proposed AO based algorithm can be proved as in Section \ref{convergence1}.
	\end{remark}
	\subsection{Complexity Analysis} The complexity of solving the SDP problem is ${\cal{O}}\left( {{m_{sdp}}n_{sdp}^{3.5} + m_{sdp}^2n_{sdp}^{2.5} + m_{sdp}^3n_{sdp}^{0.5}} \right)$, where $ m_{sdp} $ is the number of semi-definite constraints and $ n_{sdp} $ denotes the dimension of the associated semi-definite cone. For problem  (\ref{p46}), we have $m_{sdp} = 8$, $n_{sdp} = N_t^2 $. Similar to  Section \ref{complexity1}, the complexity of obtaining the passive BF is $ O_{pb} = {\cal{O}}\left(I^{\max}_{pb}\max\left(M, (2N_c)^4\sqrt{M}\log_2\frac{1}{\tilde \delta_{pb}}\right)\right) $, where $ I^{\max}_{pb} $ is the number of iterations and $ \delta_{pb} $ is the solution accuracy.
	The complexity of optimizing the deployment location is $ O_s =  {\cal{O}}\left(2N+4\right)^{3.5} $. The total complexity is $
	O\left[ {{I_{\max }}\left( {8N_t^7 + 64N_t^5 + 512{N_t}} \right.} \right.\\
	\left. {\left. { + {O_{pb}} + {O_s}} \right)} \right]
	$, while $I_{max}$ is the number of iterations in the AO based algorithm.
	
	\section{Numerical results}
	In this section, numerical results are provided for characterizing the effectiveness of our proposed designs.
	The simulation parameters of the beamformer-based and the cluster-based systems are given in Table \ref{simulation}.
	\begin{table}[htpb] \footnotesize
		\centering
		\setlength{\abovecaptionskip}{0pt}%
		\setlength{\belowcaptionskip}{0pt}%
		\caption {\textsc{Simulation Parameters}}
		\begin{tabular}{llll}
			\toprule
			&Beamformer-based NOMA&Cluster-based NOMA\\
			\midrule
			Number of users $  N  $& 2 & 4\\
			Number of clusters $N_c$ & -& 2\\
			Path-loss exponent $\alpha$ & 2.5 & 2.5\\
			$M_h$ & 5 & 5\\
			Rician factor $\beta$ & 3dB & 3dB \\
			Minimal required rate & 1 bps/HZ & 1 bps/HZ\\
			Noise power & -90dBm& -90dBm\\
			BS's location & [0,0,4]$^T$& [0,0,4]$^T$\\
			Users' location& [10,8,2]$^T$, [15,0,2]$^T$& [11,8,2]$^T$, [13,8,2]$^T$, [16,0,2]$^T$, [18,0,2]$^T$\\
			\bottomrule
		\end{tabular}
		\label{simulation}
	\end{table}
	\vspace{-20pt}
	\subsection {Beamformer-based NOMA}
	\label{NR1}
\begin{figure}[htpb]
		\centering
		\includegraphics[width=3in]{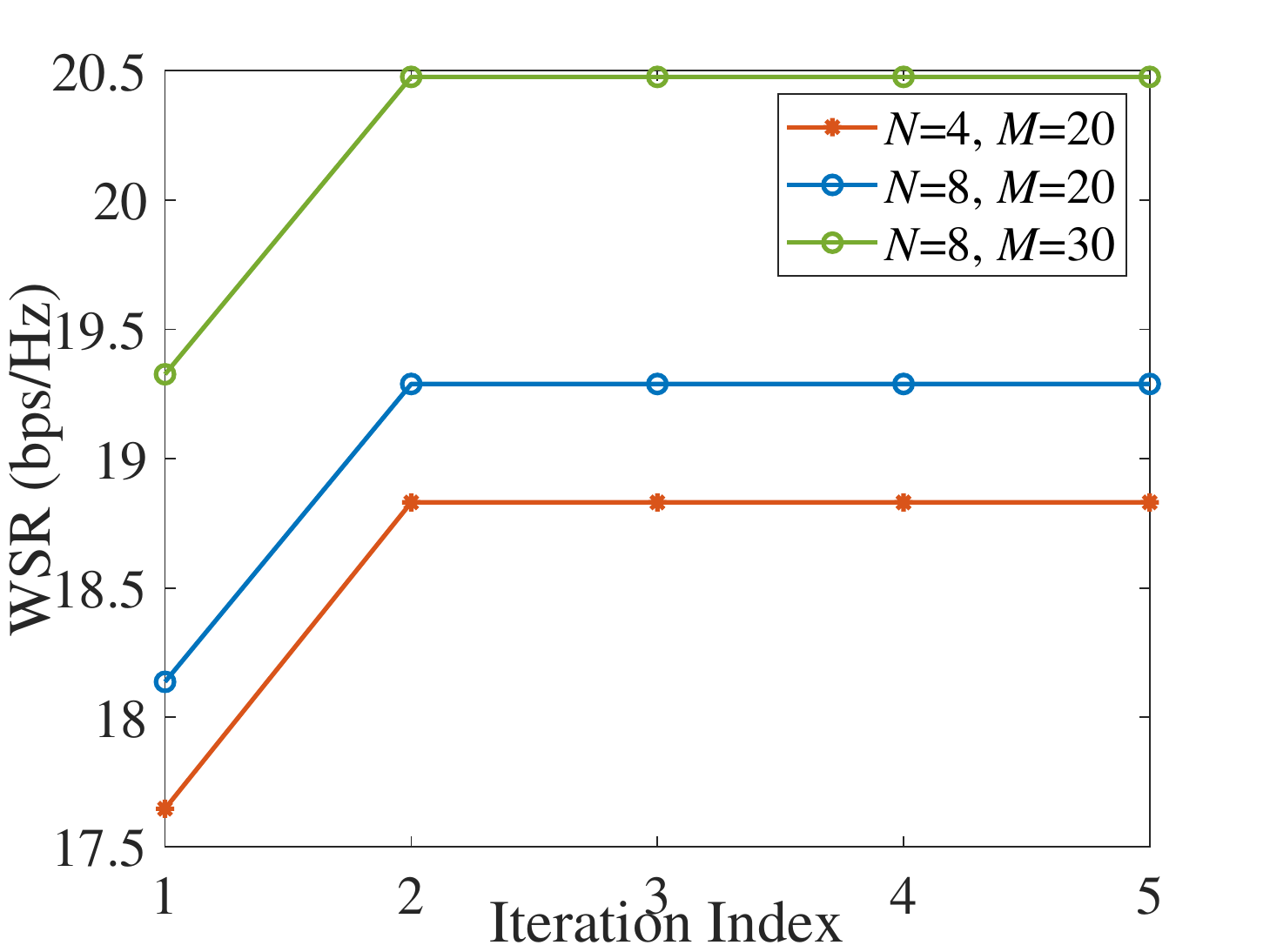}
		\caption{Convergence of the AO based algorithm for beamformer-based NOMA ($P_{\max}=30$dBm, $\eta = $[0.5,0.5]).}
		\label{sub-first}
	\end{figure}

	The convergence of our AO based algorithm proposed for the beamformer-based system is plotted in Fig. \ref{sub-first}. The maximum transmit power is set to $P_{\max}=$30dBm. The final solution is obtained by alternately solving each sub-problem, and the obtained solution for each sub-algorithm is used as the input of other algorithms. Our simulation results illustrate that the proposed algorithm converges within 3 iterations, which is consistent with {\bf{Remark 3}}.                                                                                                      
	
\begin{figure}[htbp]
		\centering
		\begin{minipage}[t]{0.49\textwidth}
			\centering
			\includegraphics[width=0.95\linewidth]{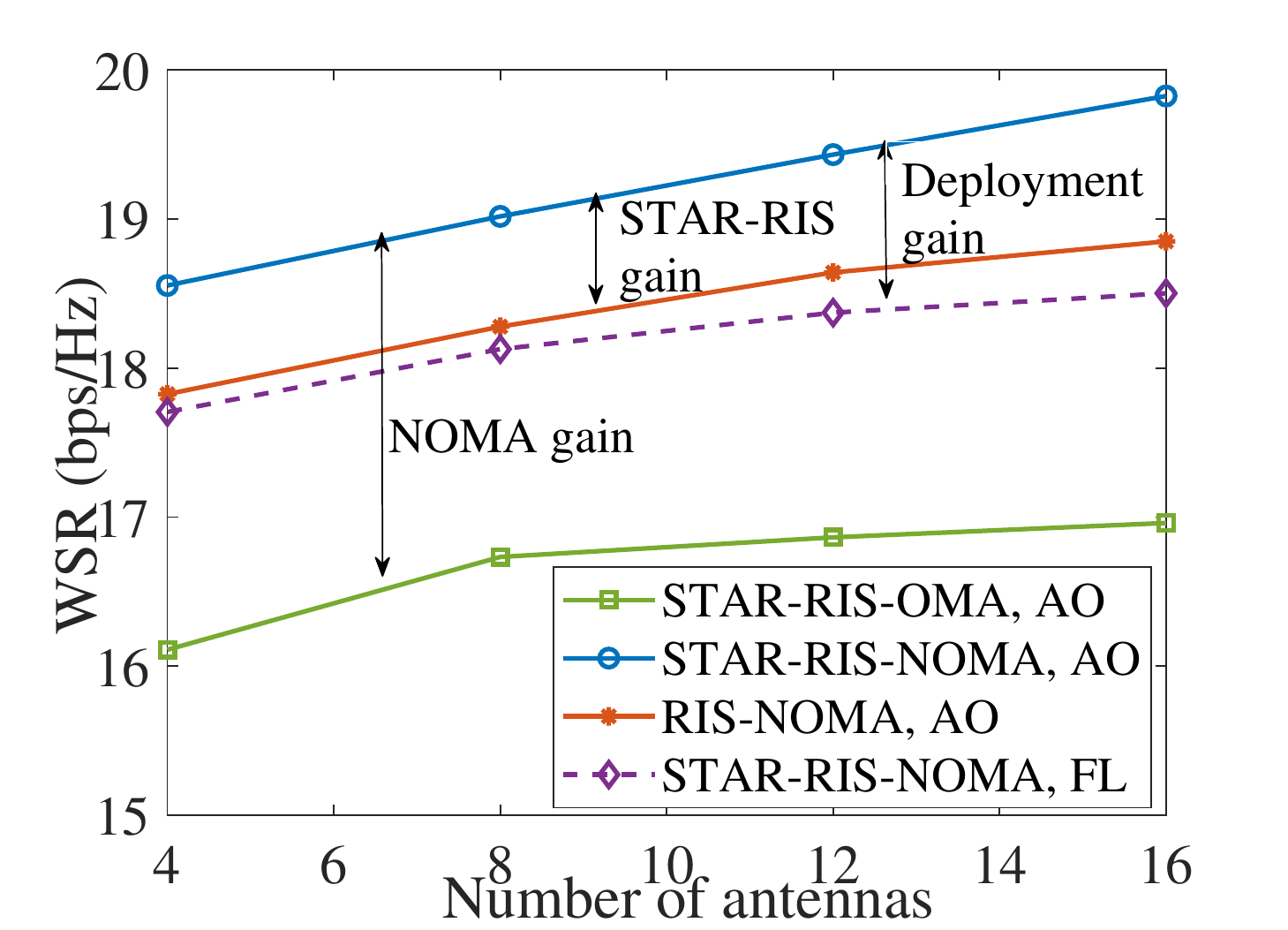}
			\caption{WSR versus the number of antennas $N_t$.}
			\label{fig3:sub-first}
		\end{minipage}
		\begin{minipage}[t]{0.49\textwidth}
			\centering
			\includegraphics[width=0.9\linewidth]{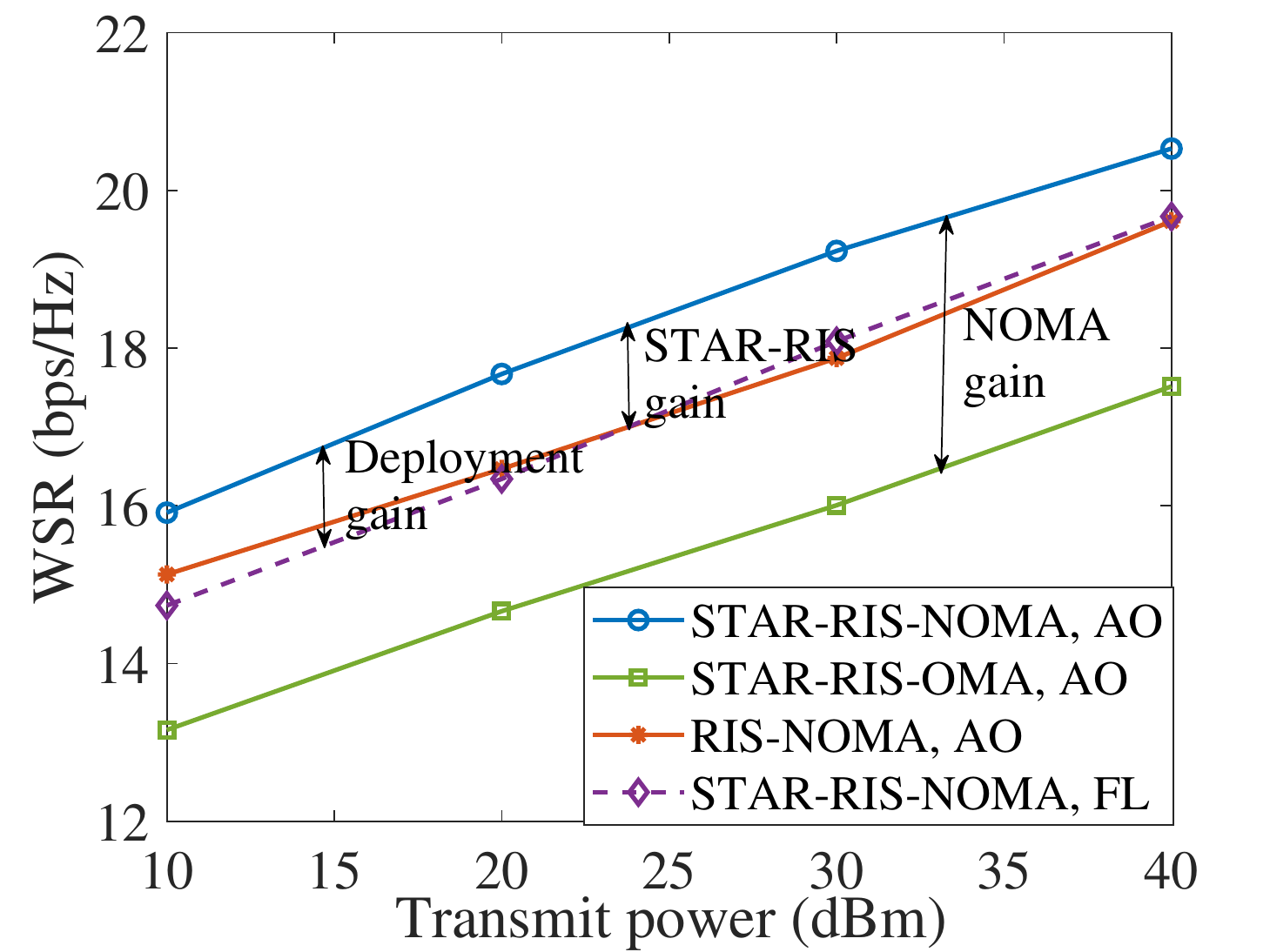}
			\caption{WSR versus transmit power $P_{\max}$.}
			\label{fig4:sub-first}
		\end{minipage}
	\end{figure}

	Fig. \ref{fig3:sub-first} depicts the WSR of the beamformer-based NOMA versus the number of antennas at the BS,  where the number of STAR-RIS elements is set to $M=$ 20, the maximal transmit power is set to $P_{\max}=$30dBm, and the users' weight $\eta$ is set to [0.5,0.5]. As seen from Fig. \ref{fig4:sub-first}, the WSR increases with $N_t$. This is expected since the increasing number of antennas leads to a higher BF gain, which improves the signal power. Regarding the performance comparison of STAR-RIS and RIS assisted beamformer-based NOMA, we can observe that STAR-RIS always outperforms the conventional reflecting-only RIS. This trend can be readily explained, because compared to the reflecting-only RIS which can only support users on the same side, STAR-RIS is capable of supporting all users due to its simultaneous transmission and reflection capability. Furthermore, NOMA outperforms OMA in STAR-RIS assisted systems, and the performance gain of NOMA over OMA becomes more pronounced upon increasing the number of antennas. Furthermore, the performance gap between the optimized and fixed location (FL) of the STAR-RIS highlights the significance of STAR-RIS location optimization.

	Fig. \ref{fig4:sub-first} shows the WSR versus the maximal transmit power of the BS, where the number of BS antennas is set to $ N_t= $ 8, the number of STAR-RIS elements is set to $ M= $ 20, and the users' weight is set to $\eta = $[0.5,0.5]. Observe that the WSR of all schemes linearly increases as $P_{\max}$ increases. Similar to Fig. \ref{fig4:sub-first}, our proposed STAR-RIS-NOMA has the best performance for the following reasons: 1) NOMA serves multiple users within the same time- and frequency resource block, hence leads to the higher spectrum efficiency; 2) the employment of STAR-RIS achieves full-space coverage and provides additional DoFs for optimization. Furthermore, the fixed STAR-RIS location leads to a degraded performance compared to our proposed AO based algorithm.
	\begin{figure}[htpb]
		\centering
		\includegraphics[width=3.2in]{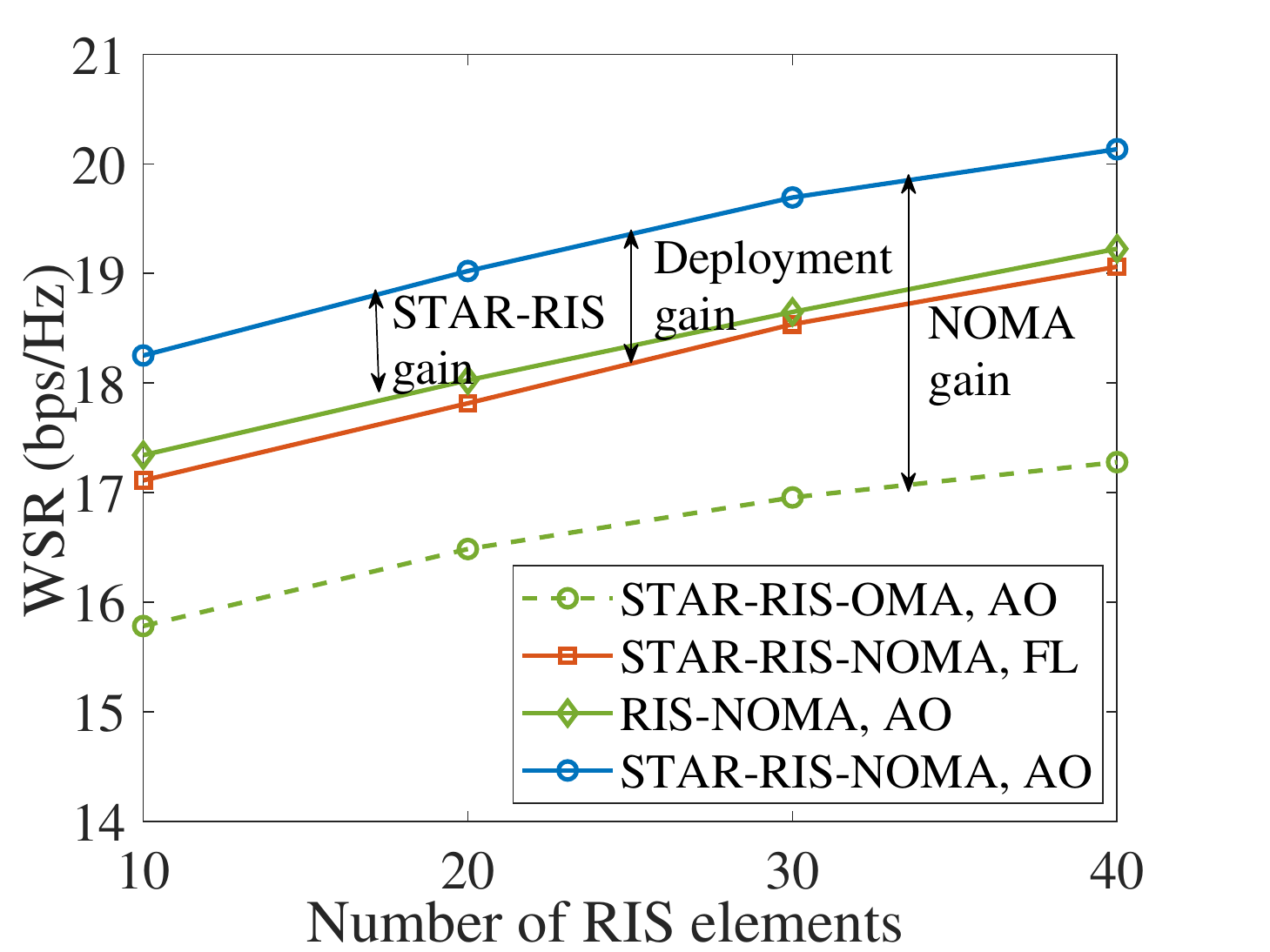}
		\caption{WSR versus the number of RIS elements $M$.}
		\label{fig5}
	\end{figure}
	
	In Fig. \ref{fig5}, we plot the WSR versus the number of RIS elements $M$, where the number of antennas is set to $ N_t= $ 8, the transmit power is set to $P_{\max}=$30dBm, and $\eta = $ [0.5,0.5]. The WSR increases upon increasing the number of RIS elements, which can be readily explained: the element-wise amplitude control yields increased passive array gains and extra DoFs for signal power enhancement and inter-user interference mitigation. Furthermore, a considerable performance loss is observed between our proposed AO based algorithm and its fixed location counterpart, which underscores the importance of STAR-RIS location design.

\begin{figure}
		\centering
		\includegraphics[width=3in]{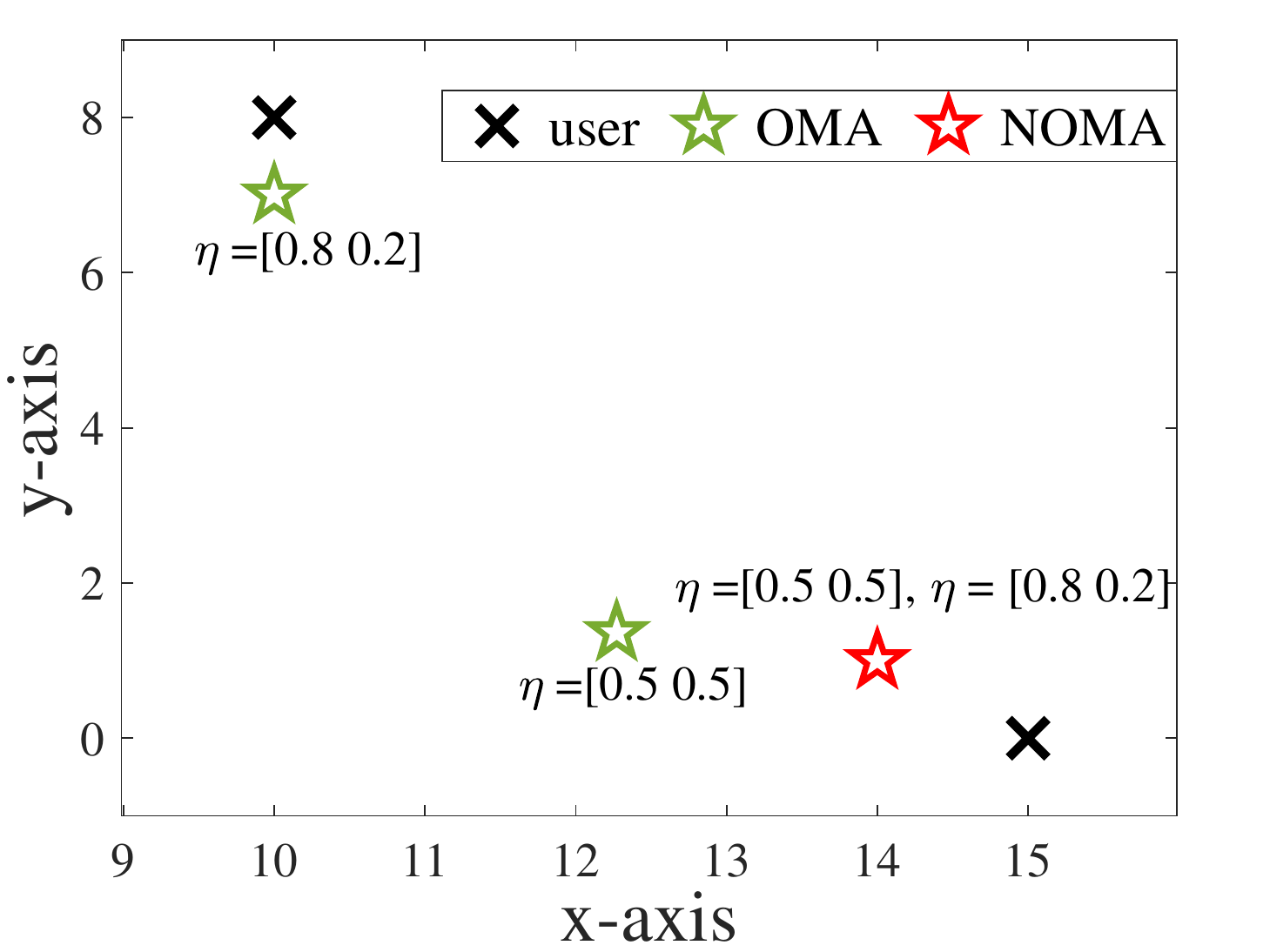}
		\caption{STAR-RIS deployment location for beamformer-based NOMA and OMA.}
		\label{deployment-beamformer}
	\end{figure}

	Fig. \ref{deployment-beamformer} shows the optimized location of the STAR-RIS of both the OMA and of the beamformer-based NOMA schemes. The deployment region of the STAR-RIS is set to $x \in $ [10,14], $y\in $ [1,7], and the initial location of the STAR-RIS is set to [12,2,2]. For the NOMA scheme, we assume a fixed decoding order, where the transmission based user is detected before the reflection based user. The optimized STAR-RIS location of NOMA for $\eta=$ [0.8,0.2] and $\eta = $ [0.5,0.5] is almost the same. It can be readily explained: once the decoding order is defined, the STAR-RIS is deployed to guarantee the higher channel gain of the reflection based user, despite its low weight. As for the OMA scheme, the weights of users have a significant impact on the STAR-RIS deployment location: the STAR-RIS should be located near the high weight user for a higher WSR.

	\subsection {Cluster-based NOMA}
	We provided numerical results for the cluster-based system in this part. The performance enhancement of our proposed algorithm is demonstrated by employing the following benchmarks for comparisons.
	
	\begin{itemize}
		\item {\bf{Optimized active BF and power allocation (OBP):}} In this case, only the active BF at the BS and the PAFs are optimized, while the passive BF of STAR-RIS is randomly generated and the location of the STAR-RIS is fixed.
		\item {\bf{ZF-RTARVs-FL:}} The active BF at the BS is obtained by the zero-forcing (ZF) method, while the passive BF is randomly generated and the location of the STAR-RIS is fixed.
	\end{itemize}

\begin{figure}
		\centering
		\includegraphics[width=3in]{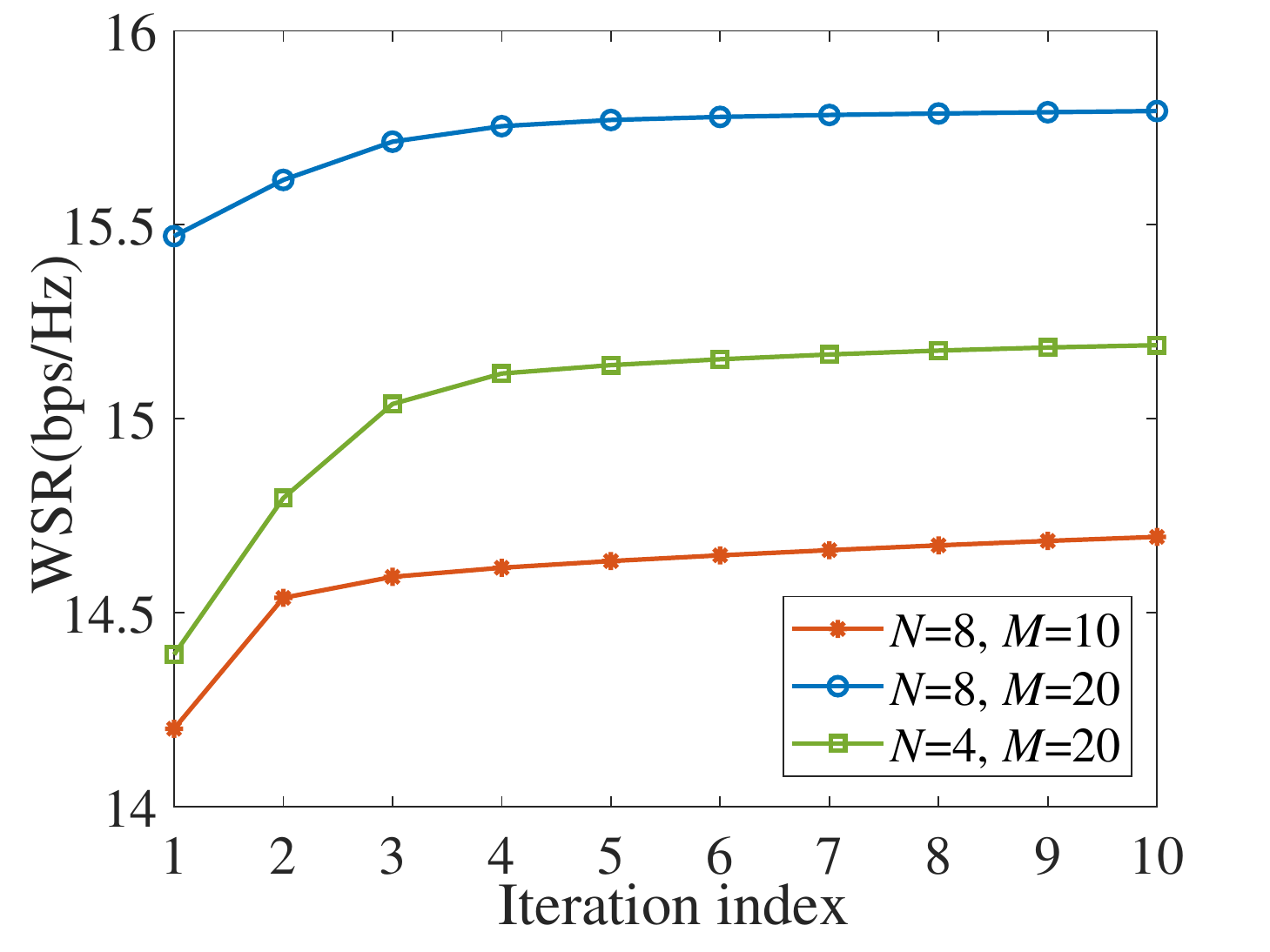}
		\caption{Convergence of our proposed algorithm for cluster-based NOMA ($P_{\max}= $30dBm, $ \eta =$ [0.25,0.25,0.25,0.25]).}
		\label{fig2:sub-second}
	\end{figure}

	In Fig. \ref{fig2:sub-second}, the convergence behaviour of our proposed AO based algorithm is investigated, where we set $P_{\max}= $30dBm and $ \eta = $ [0.25,0.25,0.25,0.25]. Results are obtained for a single random channel realization. Our results show that the proposed algorithm converges within 5 iterations. The algorithm of the cluster-based NOMA converges slower than the beamformer-based strategy. This is because the cluster-based NOMA involves more optimization variables and constraints due to the power allocation optimization in clusters.

\begin{figure}[ht]
		\begin{minipage}{.32\textwidth}
			\centering
			\includegraphics[width=1\linewidth]{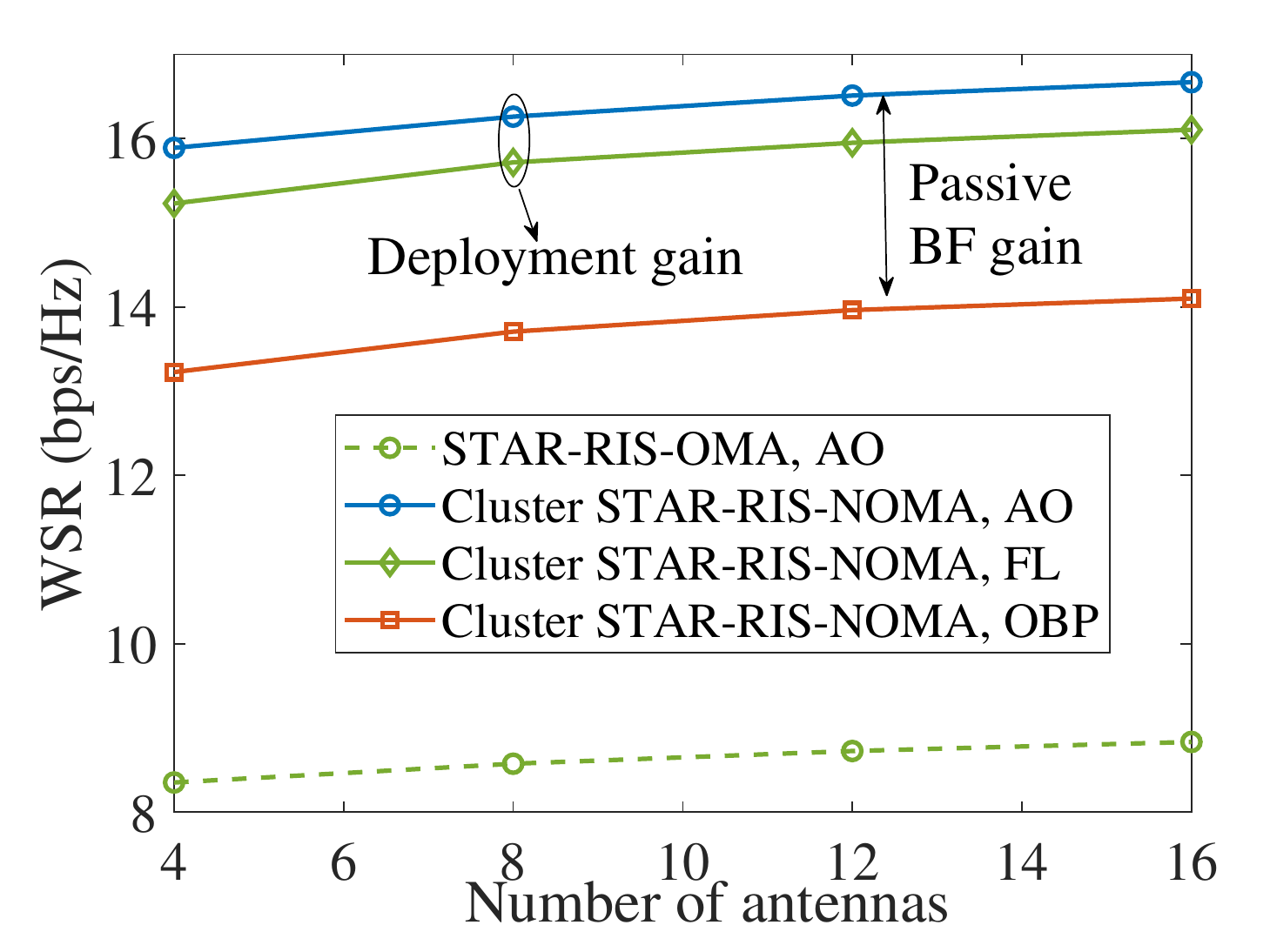}
			\caption{WSR versus the number of antennas.}
			\label{fig4:sub-second}
		\end{minipage}
		\begin{minipage}{.32\textwidth}
			\centering
			\includegraphics[width=1\linewidth]{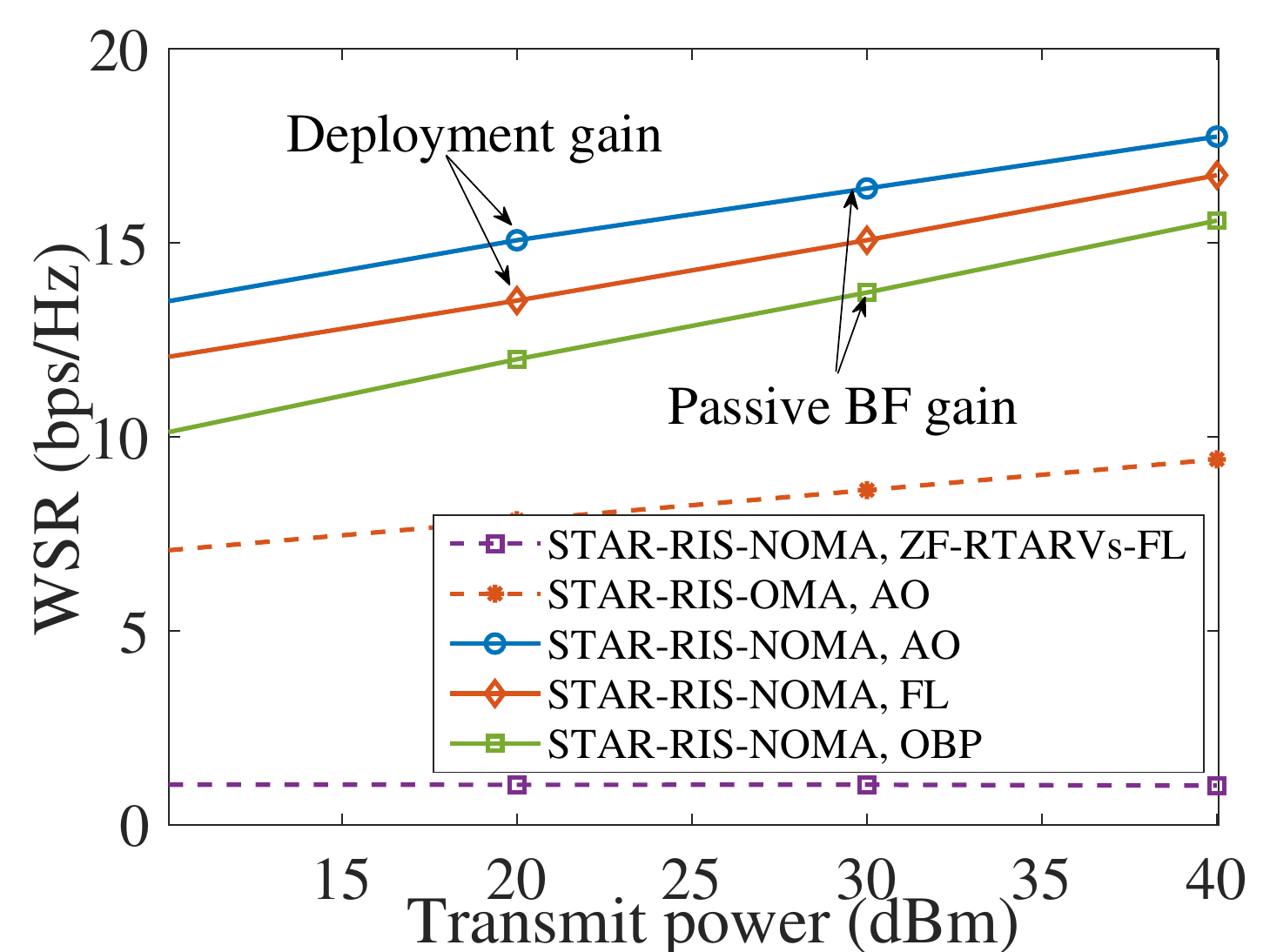}
			\caption{WSR versus the transmit power.}
			\label{fig5:sub-second}
		\end{minipage}
		\begin{minipage}{.32\textwidth}
			\centering
			\includegraphics[width=0.96\linewidth]{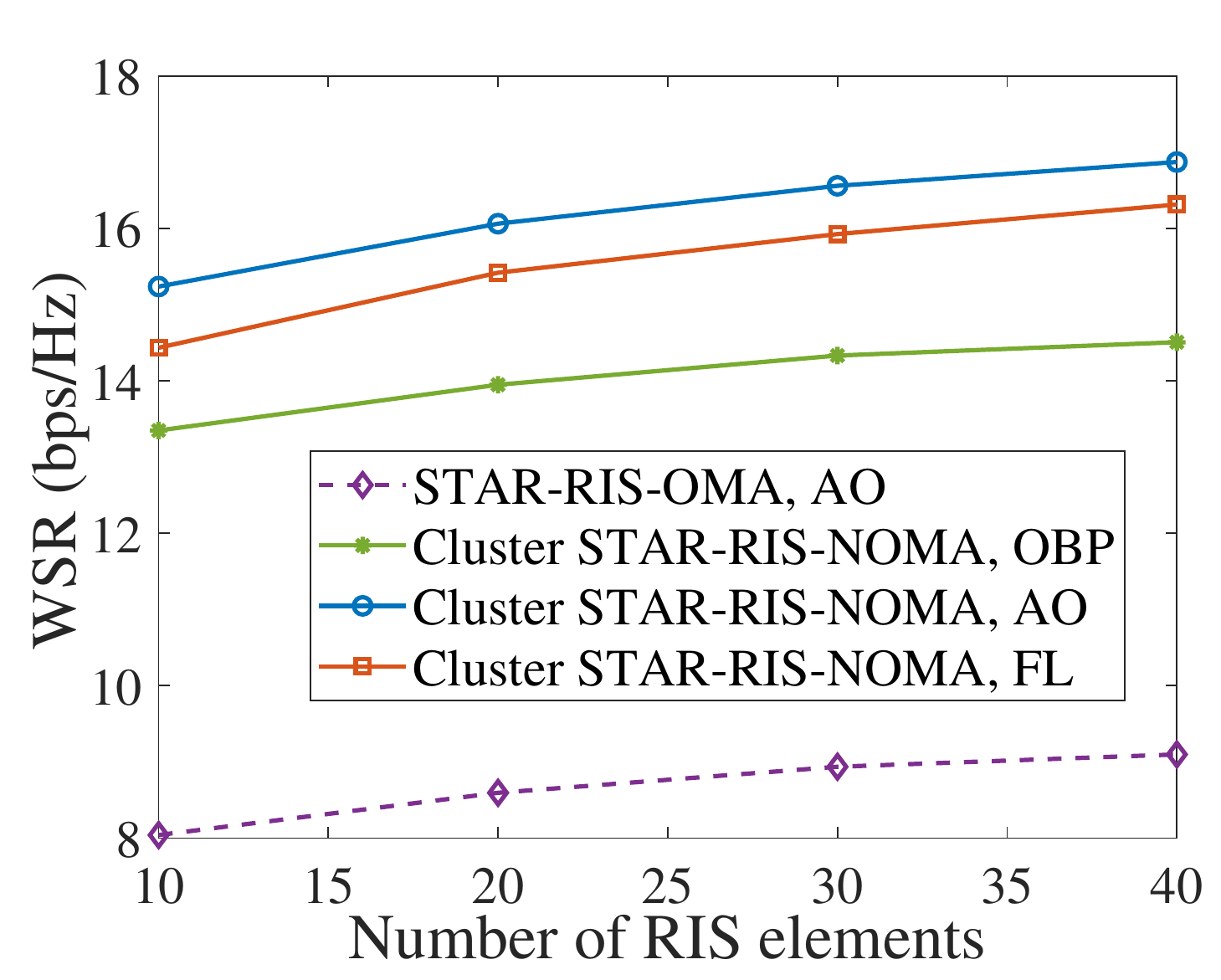}
			\caption{WSR versus the number of RIS elements.}
			\label{fig6:sub-second}
		\end{minipage}
	\end{figure}

	We plot the WSR versus the number of antennas $N_t$, the maximal transmit power $P_{\max}$, and the number of RIS elements $M$ in Fig. \ref{fig4:sub-second}, Fig. \ref{fig5:sub-second}, and Fig. \ref{fig6:sub-second}, respectively. The minimal required rate is set to $\gamma=1$ and the weight vector is set to $\eta = $ [0.25,0.25,0.25,0.25]. The WSR increases as $N_t$, $P_{\max}$ and $M$ increase, and the performance gain of NOMA over OMA becomes more pronounced for the higher transmit power. Furthermore, the proposed AO based algorithm outperforms the OBP and FL benchmarks in terms of the WSR, which highlights the significance of passive BF and deployment location optimization.
	
\begin{figure}[htpb]
		\centering
		\includegraphics[width=3in]{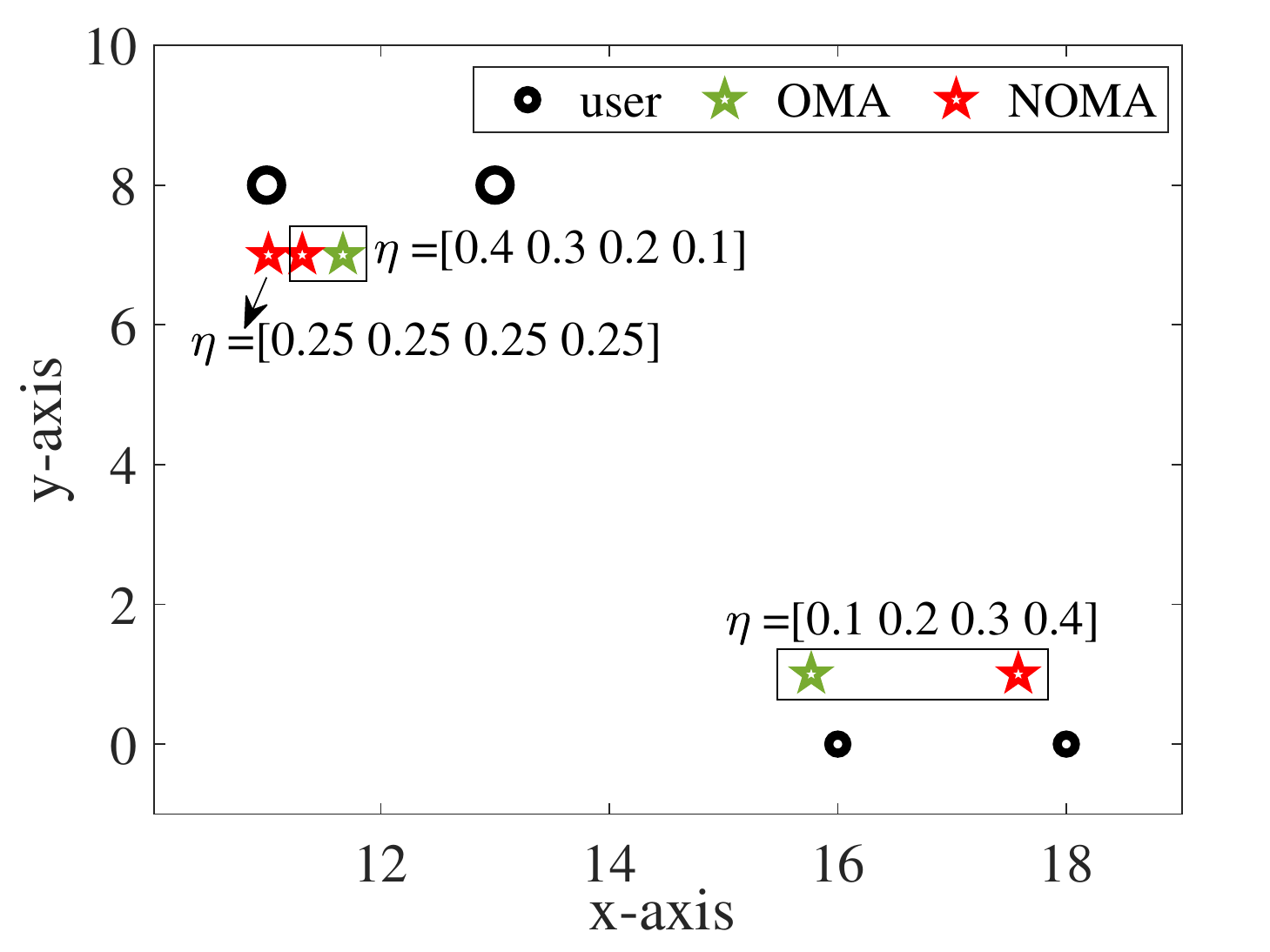}
		\caption {Optimized STAR-RIS location for cluster-based NOMA and OMA.}
		\label{deployment}
	\end{figure}

	Fig. \ref{deployment} portrays the deployment location obtained by our proposed algorithms for cluster-based NOMA and OMA schemes. The predefined STAR-RIS location region is $x \in$ [10,18], $y \in$ [1,7], and the initial STAR-RIS location is [14,3,2]. Observe that the STAR-RIS is located near the cluster having higher weights, which is in contrast to the beamformer-based NOMA. This phenomenon can be explained by the principle of cluster-based NOMA, where the SIC is only adopted among users within a specific cluster, and the path-loss of users in a cluster is approximately the same due to their similar locations. Hence the location of the STAR-RIS has less of an impact on their channel power as well as on their SIC order. Therefore, the STAR-RIS is only deployed to increase the channel gains of the cluster having high weights, regardless of their decoding order. More particularly, for the case of $\eta =$ [0.25,0.25,0.25,0.25], the STAR-RIS is deployed closer to the specific cluster, which has the highest channel gain. Furthermore, it is interesting to find that the STAR-RIS deployment strategy for OMA is more similar among users than for NOMA, regardless of the weight, which provides a useful guideline for location design.
	
	\vspace{-20pt}
	\section{Conclusion}
	In this paper, the joint location and beamforming design of the STAR-RIS assisted downlink of networks was investigated, where the beamformer-based and cluster-based NOMA were employed as multiple access schemes. Both the STAR-RIS deployment location, as well as the passive and active BF were jointly designed for maximizing the WSR of both strategies. To solve the resultant non-convex problems, we decomposed them into several sub-problems, where the SCA and SDP methods were conceived for converting these intractable non-convex sub-problems into equivalent convex ones. Then, AO based algorithms were proposed for alternately obtaining solutions of the original problems.
	Our simulation results validated the performance enhancement for employing NOMA and STAR-RIS compared to the reflection-only RIS assisted networks and STAR-RIS-OMA. More particularly, the performance enhancement attained by optimizing the STAR-RIS location was demonstrated, and our results revealed that different multiple access techniques prefer different location strategies, and that the users' weight is one of main influencing factors of the STAR-RIS's location optimization.
	
	\numberwithin{equation}{section}

	\bibliographystyle{ieeetran}
	\bibliography{IEEEabrv,RIS}

\end{document}